\documentclass[aps,nofootinbib,prd,superscriptaddress]{revtex4-1}

\usepackage{graphicx,color}
\usepackage{subfig}
\usepackage[english]{babel}
\usepackage{amsmath} 
\usepackage{verbatim} 
\usepackage{mathrsfs}
\usepackage{amsfonts}
\usepackage{hyperref}
\newcommand{\kets}[1]{\left\vert #1 \right\rangle}
\newcommand{\bras}[1]{\left\langle #1 \right\vert}
\newcommand{\beq}{\begin{equation}}
\newcommand{\eeq}{\end{equation}}
\newcommand{\bqr}{\begin{eqnarray}}
\newcommand{\eqr}{\end{eqnarray}}
\newcommand{\beql}[1]{\begin{equation} \label{#1}}

\begin{document}

\author{Sujoy K. Modak}
\email{sujoy.kumar@correo.nucleares.unam.mx}
\author{Leonardo Ort\'{i}z}
\email{leonardo.ortiz@correo.nucleares.unam.mx}
\affiliation{Instituto de Ciencias Nucleares, Universidad Nacional Aut\'onoma de M\'exico, M\'exico D.F. 04510, M\'exico}
\author{Igor Pe\~na}
\email{igor.pena@uacm.edu.mx}
\affiliation{Instituto de Ciencias Nucleares, Universidad Nacional Aut\'onoma de M\'exico, M\'exico D.F. 04510, M\'exico}
\affiliation{Plantel Casa Libertad, Universidad Aut\'{o}noma de la Ciudad de M\'{e}xico, Calzada Ermita Iztapalapa 4163, Distrito Federal, 09620, M\'{e}xico}
\author{Daniel Sudarsky}
\email{sudarsky@nucleares.unam.mx}
\affiliation{Instituto de Ciencias Nucleares, Universidad Nacional Aut\'onoma de M\'exico, M\'exico D.F. 04510, M\'exico}

\title{Black Hole Evaporation: Information Loss But No Paradox}


\begin{abstract}
The process of  black hole  evaporation resulting from the Hawking effect has  generated  an intense controversy  regarding  its potential conflict  with   quantum mechanics'  unitary   evolution.  A recent set  of   works by a collaboration involving one of us, have revised the  controversy  with the  aims of,   on  one  hand,  clarifying some  conceptual  issues  surrounding it, and,  at the  same  time,  arguing   that collapse theories  have the potential to  offer a  satisfactory resolution of the so-called paradox.  Here we show  an  explicit calculation supporting this claim  using a simplified  model  of black hole   creation and  evaporation, known as  the CGHS  model,  together with a dynamical reduction  theory, known as  CSL, and  some speculative,  but seemingly   natural  ideas  about   the  role of quantum gravity in connection with the   would-be singularity.  This  work represents a  specific realization of general ideas  first   discussed  in  \cite{Okon2} and a  complete  and detailed  analysis of a  model    first  considered   in  \cite{mops}.
\end{abstract}

\maketitle

\tableofcontents


\section{Introduction}\label{intro}

The  surprising discovery  of  black hole  radiation  by  S. Hawking \cite{hawk} in the  1970's,  has had  an enormous  influence in  our   ideas  concerning   the interface of quantum theory and
gravitation.  For instance, it  has  changed  our perception  regarding the  laws of black hole thermodynamics,  which,  before that discovery,   could have been regarded  as  mere
analogies  to  our current view that  they represent  simply the ordinary thermodynamical  laws,    as they apply to situations involving  black  holes (see for instance\cite{Wald}).    This,   in turn, has  led
to the  quest  to  understand,   on  statistical mechanical  terms,  and  within   different proposals  for a theory of  quantum gravity,  the  area of the  black hole horizon  as  a measure of the
black hole   entropy.   In fact, the most popular  programs  in this   regard, String Theory  and Loop  Quantum Gravity,  have  important success  in this front.  Furthermore,   the   fact that  as the  black hole  radiates it must lose mass  leads  to  some   tension between the   picture that  emerges  from  the gravitational side and   the basic  tenants of quantum theory. This
tension was first pointed out  by Hawking \cite{hawk2} and has even  been  described   by  many theorists  as the ``Black Hole  Information Paradox" (BHIP).
 The root of  the   tension  is that,  according to the   picture that  emerges  from the   gravitational side,  it seems  that one can start  with a pure initial quantum state characterizing the system
 at some initial stage, which  then  evolves  into something that, at the quantum level,  can only be characterized  as a  highly  mixed  quantum state,  while,  the standard  quantum
 mechanical considerations  would lead one  to expect a fully unitary evolution.
 There is  even some  debate   as to whether or not  this issue   should be considered  as  paradoxical. This  question  has  been discussed  in  \cite{Okon-New} showing   that,   at    the basis   of this  debate,   there are  some  basic  differences of outlook  about issues   such  as  the fate  of the  singularity  in the context of   a quantum  theory  that has  been  extended to cover  the gravitational realm (\emph{i.e.}, in  what sense  will the singularity   be resolved   by  quantum gravity, if at all?), and     also,  to  a  certain extent,  what should be   expected   from   quantum theory  in general (\emph{i.e.}  should   it be viewed    as  describing  a ``reality  out there",  or   as just encoding the   information   we  have  about a   given system?).

One point of view  is that  the singularity  represents an  additional    boundary of space-time  (besides   the   standard  one  associated  with  the   asymptotic  regions)   and
 thus the discussion about information,  or about the nature  of the end  state,   is erroneous if one   does not take into  account the    information   codified in such  boundary.
 This   point is,   of course,   completely justified   when  considering the situation from the purely classical view of  space-time. If, however,  as it is often expected,    quantum gravity resolves
 the singularity  replacing  it  by  a  region that can only be described  in  the appropriate quantum gravity terms,   one   would  be  justified in seeking a  clarification of the situation without  invoking an additional boundary.  Moreover,  given that such   quantum gravity region  will  leave no trace,    as  far  as  asymptotic observers  are concerned, in the  case  of complete
 evaporation of the black hole,  one  might  want to obtain an effective    characterization  of the evolution,  that  corresponds to  what is, in principle, accessible to them.  For  a more extensive
discussion of these  issues   we refer the reader to  the  work \cite{Okon-New}.

Leaving aside  these  issues,  a large number of researchers have  been  searching for  a  scheme  to address  the   black hole information problem, within the context  of   various   existing  proposals  for  a theory of quantum gravity.  This  is    natural   given the fact  that,  the ``paradox" truly  emerges only  after  one   assumes that a quantum theory of gravity   removes the  singularities  that   appear in association with black holes  in   General  Relativity, and  thus, when  contemplating any such  proposal ,  the   BHIP   issue  acquires a  new urgency.
In fact,  within the  community   that  follows the most popular approaches  to quantum  gravity,  the   subject has  recently been the focus of  intense attention.

For those  researchers  arriving at the  issue  from the String Theory perspective,    the  importance   of the  issue   is intensified  by the AdS/CFT  conjecture \cite{AdS-CFT},  which  indicates that  a theory  including   gravitation  on   the  bulk  should be equivalent to a  another theory   involving no gravitation on the   boundary,  and,  as  such,  the description   of    the   formation  and  evaporation of a black   hole
  should  be equivalent to  the  description of a process  involving no black holes. Thus,  if the  evolution is   unitary  in the no  black hole  situation, there   should be no breakdown  of unitarity in the case involving a  black hole  creation and evaporation. Thus  information  cannot possibly  be  lost  when a  black hole forms  and  subsequently evaporates via Hawking effect \cite{maldacena}.

These  arguments have  led  some physicists to argue  that  the AdS/CFT duality  implies  that information must be preserved {\it always}. Although there is no clear indication that this duality will hold for asymptotically flat spacetimes, it is conjectured  to be valid   for   situations  involving the anti-de Sitter/de Sitter/asymptotically Lifshitz space-times and conformal field theories \cite{AdS-CFT, Strominger:2001pn, Kachru:2008yh}. Moreover,  within that context,   there seems to be  no clear   explanation  of how the information is recovered in a black hole evaporation  within the space-time description,   or where  precisely  does   Hawking's analysis indicating   that  the   final state  is  not pure,  actually  go wrong. These issues are worthwhile  revisiting given  recent arguments \cite{FireWalls} indicating that  the three following
  well known physical principles cannot be satisfied simultaneously in   the context  of black hole evaporation:
\begin{enumerate}
\item Hawking radiation is in a pure state, \emph{i.e.}, the evolution of  a quantum field state is unitary and there is no loss of information.
\item The Effective Field Theory (EFT) approach  based on the  notion that, although there is a breakdown of physics at some point inside the horizon, EFT should be  well defined  and a good  description  of physics  outside the horizon.
\item The validity of the equivalence principle at the horizon, \emph{i.e.}, the infalling observers feel nothing unusual at the event horizon.
\end{enumerate}
 One finds  in the literature  various  approaches  to deal  with  the tension among  three  principles  above. For  instance,   the proposal  in  \cite{FireWalls}  prioritizes (1) and (2) over (3).  A  consequence of  such  choice   is that   the event horizon would be   turned  into a  so called ``firewall",   which represents a fundamental inconsistency   with    basic  ideas  of general relativity   embodied   in   the equivalence   principle which  would  indicate that, from the local perspective, nothing  unusual can be taking place  at the   event horizon, which after all  is  only defined  globally. The appearance of the firewall   would indicate that the event horizon is  in a sense the ``end point'' of the space-time manifold,  contrary to the basic views of general relativity, in this regard.  Other 
 proposals    consider  some  rather  exotic ideas, for example,   that the outgoing and infalling particles are connected by a worm-holes, and therefore they are not independent objects \cite{wormhole},  the  existence of Planck stars \cite{planck-star}, and the modeling   of the black hole interior with ``fuzzballs" \cite{Mathur1, Mathur2}.

  On the  Loop  Quantum Gravity  (LQG)    side,    it has   been argued that,  as the theory seems to be able to resolve the singularities,   of both ,  cosmological (see for instance \cite{LQG-singularity}) and  black hole (see for instance \cite{bh-singularity})  kind. In particular   as  in that  theory  there  would be no room for  divergences of  an energy-momentum tensor,  there   can be no firewalls,
  and  nevertheless, the information  would be leaked at late times  in the form  of  unusual  quantum  correlations.

  We must point out, however,  that,   as the  theory  of LQG  is  meant  to   involve  a    resolution  of   the    black hole  singularity, the corresponding   region (\emph{i.e.} the region  which,  in the classical  characterization,  would   contain  the  singularity)   would have to   be  a  region   with  exotic  properties,  where,     in all  likelihood,  the   ordinary space-time notions    would cease to  be  valid. Therefore,  it is not  completely clear how exactly, the information  would traverse  across  such    exotic  region: In fact,  in the  2  dimensional  example  based on the CGHS  model  presented in the  work  \cite{Ashtekar:2008jd}, the region  corresponding to  the ``would be singularity" is  replaced by a region where the  conformal factor  (which   characterizes  the  space-time    metric   which  is conformally  flat),    undergoes fluctuations  about zero. That is,  we  have a region where  the metric   signature fluctuates\footnote{Strictly speaking, as the  example is  two dimensional, what fluctuates  is not really the signature, but the  specific directions that are time-like  and  space-like.} and,  as  far as we know,  the evolution of  a quantum field through such a  region is  not well understood (see  \cite{Bojowlad2}  for  a recent work  discussing  such  problems  in  detail). Another issue that   does not  seem  to  be   addressed  in a satisfactory  way   in that proposal is  related to the  problems  faced in   other  attempts to solve  the information loss question:    If  most of the energy of the initial  black hole  is  emitted  during the normal  Hawking radiation state  of the  evaporation, then  there would  be  very  little energy left  to  be  radiated in the late stages,  which  are presumably those  where  QG   effects  would  be relevant. In fact,  even if QG resolves the  singularity, and information is   somehow  able to cross to the other side of the quantum gravity  region, the  amount of information  encoded  cannot   be too large. This  is  simply  due to the limitations  associated    with  the small  energy  available to populate  different   highly  localized  modes of   the   quantum field. In other words,  it seems  one   would need to  face   a  serious   energy deficit   if  one  wants to  argue that  there  are   enough  modes  excited in the  radiation which  escapes through the singularity,  so  the complete  state  of the quantum field  in $\mathscr{I}^+$ is pure,   even though  the    restriction of   the state   to the   early  part of  $\mathscr{I}^{+}$  is   both  thermal and contains most of the energy. We believe that   these   facts   cast   some   doubts   on the claims  that  the information is preserved,    and   that  the   final state  must be   unitarily   related  to the initial  one.

Another important issue that has to be stressed in connection with any  such  proposal to  dealing   with  the BHIP,  is that, among  other things,   it  should  account for the   fact  that  a pure state  must   turn  into  an ordinary (quantum) thermal state corresponding  to  a \emph{proper} mixture (rather then an \emph{improper} one, see \cite{dEspagnat} for terminology). Effectively, as the interior region of the black hole disappears when the singularity is removed, the state of the field for asymptotic observers at $\mathscr{I}^+$ has to be described by a density matrix representing an ensemble, every element of which, is in a pure state (which one ignores) and not by a density matrix that results from tracing out degrees of freedom of some region of space-time.

In our view,   there is  little hope that   these  problems could  be  completely  clarified  without  first  setting   them in a  proper context. The   fact  is  that  there  is not  even a  full consensus about how  should  we  view quantum theory   in the  absence of the  gravitational  complications. In fact,  the  so called  ``measurement problem"  (often  characterized  as the ``reality
problem")   in  quantum mechanics   remains, almost a century after the theory's  formulation,  a  major obstacle to considering  the  theory  as truly fundamental.  We  shall see  that  some
related  issues appear  in  unexpected  places  along  our discussion of  our proposed  resolution   of the BHIP.

The  search for a satisfactory   interpretation  of the theory, despite the  efforts  of   many  insightful  physicists, continues,  and all the  existing work  has  not  yet yielded  a   convincing  option,  at least not one that  is universally   accepted.

 The  basic difficulty, as  described  for instance  in \cite{measurement, Penrose},  is the  fact that the  theory,  as presented in  textbooks,   relies  on two different and incompatible  evolution  processes.   As  R.  Penrose \cite{Penrose}  has  characterized  them, the theory relies on two different evolution processes: the  $U$ (unitary)  process,  where the state  changes  smoothly  according to Schr\"odinger's   deterministic  differential  equation, and  the $R$ (reduction) process, in which the state of the system  undergoes  some   instantaneous  change or jump, in
 an un-deterministic   fashion. The  $U$ process  is  supposed to control a system's  dynamics  all the time that  the system is  not interfered  with,  while  the $R$  process takes  control
 whenever a measurement  is  carried  out.

 The   fundamental problem  is that no one has been able to characterize, in a general  way, when exactly  should  a  physical process be considered  as a {\it measurement}.  This issue has   been  studied   in   depth  and   debated,    according to most people, to exhaustion,  in the scientific and  philosophical  literature \cite{More-measurement},  with no  universally  acceptable  resolution (for extensive discussions about  dealing with  these  issues   in terms of   interpretational  proposals,   and    their related   shortcomings  see   for instance \cite{Interpretaciones}). Moreover,   the fact  that  in laboratory situations,  one clearly knows  when a measurement has been carried out  has  led  some people to  claim the debate as   irrelevant,  and  most   professional researchers now  advice  their students not to think about the issue and  just calculate.   Nonetheless, as characterized by J. Bell \cite{Bell}, this  kind of {\it for all practical purposes} (FAPP)  approach,  is  not  fully satisfactory  at the foundational level, as it involves treating  the system differently from the measuring device or the observer, and this division is one for which the theory offers no specific internal rules.

These  issues  are often dismissed by large segments of the  physics  community  as  simple  philosophical/interpretational dilemmas  with  little, if any, relevance  for the application of the  theory.  Needless  to say that  there  are other  colleagues   who  strongly disagree  with  such  characterization,  and  that, as it is  evident,   we find  ourselves in agreement  with this  latter  group.    In fact, it is  worthwhile   to note  a  relatively  recent work  \cite{Mau:95}, which  helps putting  the  issue  in  a clear perspective  by showing that there is  a fundamental  incompatibility between the  following  notions:

 a) the wave function provides a complete  characterization   of a system;

 b) the wave function always evolves  according  to a  deterministic  linear dynamical equation;

  c) measurements always have determinate outcomes.

\noindent Thus for instance, hidden variable theories negate a), objective collapse models negate b){\footnote{More precisely, objective collapse models negate the assumption  that all dynamical elements of the theory evolve according to a set of deterministic dynamical equations \cite{pearlepc}. The  subtlety   refers to the  fact  that  such models  involve stochastic  elements    entering  into  the   evolution equation of the  quantum state,  but of course,  when  those  elements   acquire specific  values, the evolution of the quantum   state   of  system  is   completely determined. However   the  specific  values of those  stochastic elements  are  not  only { \it  a priori}  unkown, but they are, in principle,   {\it  a priori} unknowable.}, while the  many-worlds scenarios negate c).

  The  logical  self consistency requirement   to abandon  one of the three desirable   items above,  clearly illustrates  the fact that,   when  dealing with issues of principle, as  we  necessarily  do when considering questions such as  the fate of information,   in light of  Hawking   evaporation of a black holes,  we  need to  consider,   with some  care,  the interpretational  aspects of quantum theory.


One approach to deal  with  this unsatisfactory  aspect  of standard quantum theory 
is to consider  modifying it     by incorporating  novel  dynamical features  that  avoid  the need to  distinguish between the $U$ and the $R$  process  (in the sense of  having to know  when to apply one or the other). That is,  the modification   incorporates   something like   ``the collapse of the wave function"  at the basic    dynamical  level, and in doing so removes  the  issue  completely.
The    exploration of  these   ideas  has  a long history, with the first   suggestions \cite{Bhom-Bub}   as  far back as the  mid  60's.  The   more   recent    developments  might be traced to  the  early works  of P. Perale \cite{Pearle:76},
followed  by the first truly viable proposal   the GRW  ( Ghirardi-Rimini- Weber)   theory  in \cite{GRW:85} \cite{GRW:86},  which  considers individual  stochastic  collapse  events   leading to  a  spontaneous  spatial  localization of  the   wave function of multiparticle  systems.   Further developments  led to  a   continuous  version \cite{Pearle:89}  known as  CSL (continuous  spontaneous localization),   and  to  an improved  GRW version\cite {GRW:90}.  The   work along  those lines has  continued  and  has  led to   important insights and   even to  the  development of an experimental program to test these ideas.  We  suggest the interested  reader to  consult the relevant and  exciting   literature.



In this work, we  will   show,  in a  simple  model,    how the incorporation of these  modifications of quantum theory,    together with a few  other   more or less natural assumptions,   could
lead to  a resolution to the paradox.

 The basic idea,  first discussed in  \cite{Okon2},  is that, as  the theories involving   dynamical  spontaneous  reduction of the wave function, do  generically  prescribe  a un-deterministic and non-unitary stochastic evolution, the loss of information occurs,  not  just  in  the context of black hole  evaporation,  but it takes place always.   Thus,    within  this  context,  the  situation involving black holes needs  not,  in principle,  be different  from   more  mundane  situations. That  is,  as the fundamental quantum theory    would now involve an actual  departure from  the simple linear  and unitary evolution, the fact that  the complete   evolution leads  us to a loss of information  is no longer paradoxical.  Information is  lost all the time in quantum theory and   the  unitary  evolution is only an approximation.
	
  However,  in order to  present a convincing  argument, we  need  to  do much   more than  just  to point to these simple conceptual changes.  Our task is to show  how the initially pure state   that   evolves to  form   a  black hole,   will  evolve in the  remote future  into a state that  is  characterized,  to  a  very  good  approximation,  by the   almost thermal state that  is inferred from the  Hawking  type calculations. At the same time,  we  must    ensure that   this  can be accommodated  within the  very  stringent bounds  on departures  from ordinary quantum  theory that   have  been  obtained  when  examining  the  phenomenological  manifestations of   these  theories \cite{Bassi:13}. In fact, as it was mentioned in \cite{Okon2}, and as we will explicitly show here, in order to account for the information loss in black holes, we must introduce a new hypothesis that the rate of such stochastic, dynamical state reduction is enhanced in a region of high spacetime curvature. As the spacetime curvature increases towards the singularity of a black hole, this modified quantum behavior erases almost all the information associated with an otherwise unitary quantum evolution.
	
We  should    note that, in the course of this    analysis,   we  would need to deal with  at   least  three  additional issues: what  does  quantum gravity  say regarding the  black hole   singularity and   what are the implications of   the   way  in  which it    presumably resolves  the singularity; what is   precisely the  nature of a   mixed   state, in general,   and  of a thermal state in particular? (\emph{i.e.}, does  it reflect   only our  ignorance, or is it   essentially attached  with   ensembles  of systems rather than a single one?   etc.);     and finally,  what   are the  specific  attributes    required  from   a theoretical proposal,  so  that    we would  consider it  to be  able to   account for the required  evolution  of    the quantum  field  in the conditions  associated  with a  black hole  evaporation?

The article is organized as follows:   In section \ref{sec:CGHS} we present a review of the geometry of the CGHS model, and in section \ref{QFT-CGHS} we review the standard analysis of QFT and Hawking radiation in this space-time, we also  present an analysis of the energy-momentum fluxes across the horizon associated to the quantum field vacuum state, and discuss its renormalizability. Section \ref{sec:thermal-states} contains a  brief  discussion  clarifying the  nature of the  two  kinds of mixed/thermal  states.  In section \ref{sec:CSL},  we review the standard CSL dynamical reduction theory, and in section \ref{sec:appl_CSL} we present our proposals for adapting CSL to the CGHS scenario, and for  what would be the role in the final evolution of a theory of Quantum Gravity. Finally, in section \ref{sec:final-result}, we present the final result of the evolution, and we end with some discussions on section \ref{sec:discussions}. There are four Appendices \ref{Hadamard}, \ref{proof}, \ref{int} and \ref{ssc} at the end of this paper. In Appendix \ref{Hadamard} we discuss the non-Hadamard behavior of certain states, in Appendix \ref{proof} we prove the anticipated behavior of the spacetime foliation, whereas, Appendix \ref{int} contains some useful expressions, and finally in Appendix \ref{ssc} a scheme to incorporate backreaction in presence of the wavefunction collapse is sketched.

Regarding notation, we will use signature $(-+++)$ for the metric,  and throughout this paper we set the unit $c=G=\hbar=1$.

\section{Brief review of the CGHS model}\label{sec:CGHS}

The Callan-Giddings-Harvey-Strominger (CGHS) model \cite{CGHS92} involving  black hole formation
has very close similarity with the spherical collapse of massless scalar field in four spacetime dimensions {\footnote{Although there are two subtle differences, namely, (i) the dilaton potentials are different in these two models,  thereby making different impacts on technical details and, (ii) the scalar field couples with the dilaton in  the spherical collapse,  whereas it is uncoupled to that in  the CGHS model.}}. 
Furthermore, it provides a consistent theory of  a toy version of quantum gravity in two spacetime dimensions coupled to conformal matter. Due to these features, the CGHS model gives a very  useful tool for getting insights on the  formation and evaporation of four dimensional black holes. In the last twenty years, this model has often been referred  toas a ``laboratory for testing general ideas on more realistic black holes" (see \cite{sGidd92}, \cite{aStr95}, \cite{Benachenhou:1994af} for reviews and the monograph \cite{aFabjNs05}). Particularly, it has been extensively studied to incorporate back reaction effects due to quantum matter fields \cite{ST, RST, apr}, building a toy model   for quantum gravity \cite{Kuchar:1996zm, Varadarajan:1997qz}, and also shading light on the information loss paradox \cite{apr, Ashtekar:2008jd}. We  will  follow   this tradition  and  use  it as a laboratory for testing our ideas  about   the resolution of the BHIP.     The  work in this  section and the  next,  is  just a  review  of existing work, and  contains nothing original (except, perhaps,  for  the  discussion  in sub-section \ref{Haddamard}). Those  readers familiar with the model can safely  proceed  directly to  section  \ref{sec:CSL}.

The action for the CGHS model \cite{CGHS92, sbGidwmNel92} is given by
\begin{equation}
S=\frac{1}{2\pi}\int d^2x\sqrt{-g}\left[e^{-2\phi}\left[R+4(\nabla \phi)^2+4\Lambda^2\right]-\frac{1}{2}\sum_{i=1}^{N}(\nabla f_{i})^2\right],
\end{equation}
where $\phi$ is the dilaton field, $\Lambda^2$ is a cosmological constant, and $f_{i}$ are $N$ matter fields. In this work we will restrict ourselves with only one field $f$.
 For  the most direct  path that leads to a black hole solution  one   investigates this model in the ``conformal gauge":
\begin{equation}
ds^2=-e^{2\rho}dx^{+}dx^{-}
\end{equation}
in null coordinates $x^{+}=x^{0}+x^{1}$, $x^{-}=x^{0}-x^{1}$. In this setting, the   field equation  for  the  scalar field  $f$ decouples, and the  most  general solution   can   be  written in the following manner
\begin{equation}
f(x^{+},x^{-})=f_{+}(x^{+})+f_{-}(x^{-}).
\end{equation}
This is a characteristic of CGHS model where left and right moving field modes do not interact with each other. Therefore, there  is not need to deal  with  any  sort of  ``back-scattering" effects. For  any given functions $f_{+}$ and $f_{-}$, one can then find solutions for $\phi$ and $\rho$ \cite{CGHS92}.  A particular case  is  the vacuum solution ($f=0$) \cite{sGidd92}
\begin{eqnarray}
ds^2 &=& -\frac{dx^{+}dx^{-}}{M/\Lambda-\Lambda^2 x^{+}x^{-}},\hspace{0.2cm}(-\infty < x^{+} < \infty,~-\infty < x^- < \infty ), \\
 e^{-2\phi} &=& \frac{M}{\Lambda}-\Lambda^2 x^{+}x^{-},
\end{eqnarray}
which corresponds to a black hole of mass $M$. This mass is the ADM mass \cite{apr}. The case  $M=0$ is known as the {\it linear dilaton vacuum}  solution. One can ``glue together" the linear dilaton vacuum and  the black hole solutions along the line $x^{+}=x_{0}^{+}$ by considering  a pulse of left moving matter with energy momentum tensor
\begin{equation}
T_{++}=\frac{1}{2}(\partial_{+}f)^2=\frac{M}{\Lambda x_{0}^{+}}\delta(x^{+}-x_{0}^{+}).
\end{equation}
This gives the solution
\begin{eqnarray}
ds^2 & = &-\frac{dx^{+}dx^{-}}{-\Lambda^2x^{+}x^{-}-(M/\Lambda x_{0}^{+})(x^{+}-x_{0}^{+})\Theta(x^{+}-x_{0}^{+})},\\\nonumber\\
    &&  (0< x^{+} < \infty,-\infty < x^- < 0 ), \nonumber \\ \nonumber\\
e^{-2\phi} & = & -\Lambda^2x^{+}x^{-}-\frac{M}{\Lambda x_{0}^{+}}(x^{+}-x_{0}^{+})\Theta(x^{+}-x_{0}^{+}).
\end{eqnarray}

\begin{figure}
\centering
\includegraphics[scale=0.5]{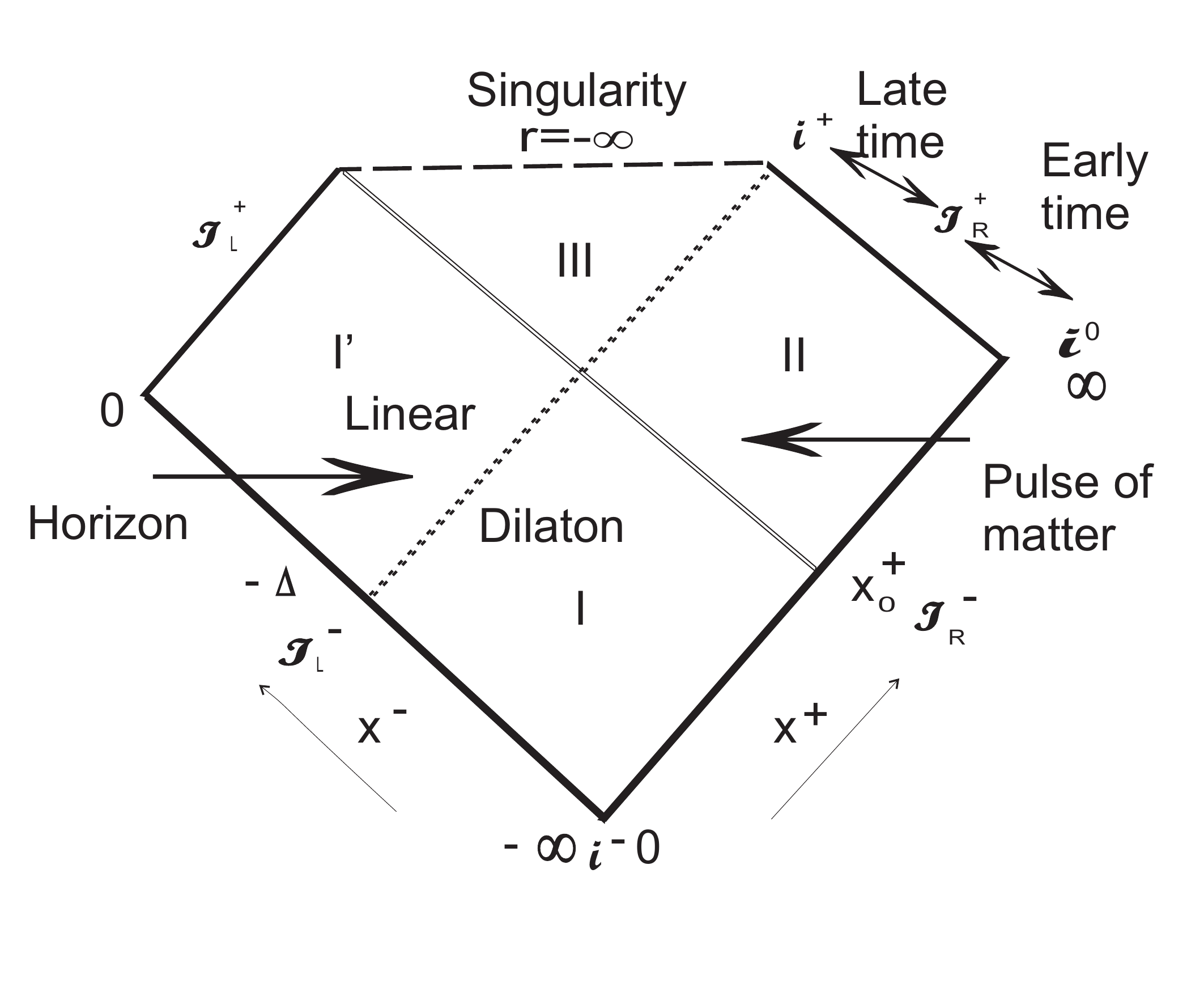}
\caption{Penrose diagram for CGHS spacetime created due to matter collapse and evaporated due to Hawking effect.}
\label{cghs}
\end{figure}

Before   (\emph{i.e.}  to the past  of)  the  matter pulse,  the space-time metric is   just  the linear dilaton vacuum solution (regions I and I' in Fig. \ref{cghs})
\begin{equation}
ds^2=-\frac{dx^{+}dx^{-}}{-\Lambda^2 x^{+}x^{-}}, (0 < x^{+}< x_{0}^{+},-\infty < x^- < 0 ),
\label{dvc}
\end{equation}
 and after $x_{0}^{+}$ it turns into a black hole solution. For later purposes it is useful to write the metric for the black hole region (regions II and III) as \cite{sbGidwmNel92}:
\begin{equation}
ds^2=-\frac{dx^{+}dx^{-}}{\frac{M}{\Lambda}-\Lambda^2 x^{+}(x^{-}+\Delta)}, (x_{0}^{+}< x^{+} < \infty,-\infty < x^- < 0 ),
\label{eq:1}
\end{equation}
where $\Delta=M/\Lambda^3x_{0}^{+}$. The position of the horizon is given by $x^{-}=-\Delta=-M/\Lambda^3x_{0}^{+}$. The Ricci curvature scalar has the form
\begin{equation}
R = \frac{4 M \Lambda}{M/\Lambda - \Lambda^2 x^+ (x^- +\Delta )}.
\label{ricci}
\end{equation}
 The position of classical singularity, where  $R$ ``blows  up",   is   given by
\begin{equation}
M= \Lambda^3 x^+ (x^- + \Delta).
\end{equation}
  We note that  the metric eq.  (\ref{eq:1}) is asymptotically flat in the black hole region $x^{+}>x_{0}^{+}$. To see this, the first step would be to use null coordinates $\sigma^{+}$ and $\sigma^{-}$, where
\begin{equation}
e^{\Lambda\sigma^{+}}=\Lambda x^{+},\hspace{0.2cm}e^{-\Lambda\sigma^{-}}=-\Lambda(x^{-}+\Delta),
\label{trans1}
\end{equation}
and $-\infty<\sigma^{\pm}<\infty$. Note that the above relationships between the coordinates $\sigma^{\pm}$ and $x^{\pm}$ are the Kruskal transformations where the latter represents null Kruskal coordinates. The $\sigma^{\pm}$ coordinates are only defined outside the event horizon (Regions I and II in Fig. \ref{cghs}). In these new coordinates the metric is given by
\begin{equation}
ds^2=-\frac{d\sigma^{+}d\sigma^{-}}{1+\Delta\Lambda e^{\Lambda\sigma^{-}}} ; ~(-\infty < \sigma^{-} < \infty, -\infty < \sigma^{+} < \sigma_{0}^{+}),
\end{equation}
if $\sigma^{+}<\sigma_{0}^{+}$ (Region-I in Fig. \ref{cghs}) and
\begin{equation}
ds^2=-\frac{d\sigma^{+}d\sigma^{-}}{1+(M/\Lambda)e^{\Lambda(\sigma^{-}-\sigma^{+)}}} ;~(-\infty < \sigma^{-} < \infty, \sigma^{+} > \sigma_{0}^{+}),
\label{sgma}
\end{equation}
if $\sigma^{+}>\sigma_{0}^{+}$ (Region-II in Fig. \ref{cghs}), where $\Lambda x_{0}^{+}=e^{\Lambda\sigma_{0}^{+}}$. In order  to exhibit the asymptotic flatness, we first write the metric in Region-II using coordinates ($t,\sigma$) defined as $\sigma^{\pm} =t \pm \sigma$ 
and then it is easy to express the metric in Schwarzschild-like coordinates using the transformation,
\begin{equation}
r= \frac{1}{2\Lambda} \ln(e^{2\Lambda\sigma} + \frac{M}{\Lambda}).
\end{equation}
The resulting metric in Schwarzschild gauge is,
\begin{eqnarray}
ds^2 = -(1-\frac{M}{\Lambda}e^{-2\Lambda r}) dt^2 + \frac{1}{(1-\frac{M}{\Lambda}e^{-2\Lambda r})} dr^2;~(-\infty < t < \infty, -\infty < r < \infty). \notag\\
\end{eqnarray}
In these coordinates the position of the horizon is given by $r_h= \frac{1}{2\Lambda}\ln(M/\Lambda)$ and singularity is situated at $r= - \infty$. Also at  spatial infinity ($r= \infty$) one has the flat metric.

In addition, one can  introduce new coordinates $y^{\pm}$ covering the whole manifold (all regions in Fig. \ref{cghs}) in the following way,
\begin{equation}
x^{+}=\frac{1}{\Lambda}e^{\Lambda y^{+}},\hspace{0.2cm}x^{-}=-\Delta e^{-\Lambda y^{-}}.
\label{ypm}
\end{equation}
These coordinates are particularly helpful to characterize the flat dilaton vacuum region eq.  (\ref{dvc}) as  there the metric  takes  the  usual  Minkwoskian form (Regions I and I' in Fig. \ref{cghs})
\begin{equation}
ds^2=-dy^{+}dy^{-}; ~ (-\infty < y^+ <  \frac{1}{\Lambda} \ln(\Lambda x_0^+) , -\infty < y^{-} < \infty) \label{iny}.
\end{equation}
On the other hand,  the black hole regions II and III in Fig. \ref{cghs} described   in terms of $y^\pm$ coordinates (using eq.  (\ref{eq:1}) and eq.  (\ref{ypm})), takes the form
\begin{equation}
ds^2=-\frac{dy^{+}dy^{-}}{e^{\Lambda y^- \left(\frac{M}{\Lambda^2 \Delta}e^{-\Lambda y^+} + e^{-\Lambda y^-} -1\right)}}; ~ (\frac{1}{\Lambda} \ln(\Lambda x_0^+) < y^+ < \infty  , -\infty < y^{-} < \infty) \label{outy}.
\end{equation}
Although, these  coordinates are not truly  Schwarzschild-like, they  are helpful for various purposes in our study.

Note that unlike the coordinates  $\sigma^{\pm}$ the  coordinates  $y^{\pm}$  cover the whole  space-time.
 The relation between these two group of coordinates, in the region II of Fig. \ref{cghs},
 is the following
\begin{eqnarray}
y^+ &=& \sigma^+ \\
y^- &=& -\frac{1}{\Lambda}\ln(1+\frac{1}{\Lambda\Delta} e^{-\Lambda\sigma^-}). \label{smy}
\end{eqnarray}
Note that the 
event horizon is located at $y^{-}=0$. We shall use these ($y^{\pm}$) coordinates in the construction of the quantum field modes in the  CGHS background (particularly in region I, I' and III in Fig. \ref{cghs}).


\section{Quantum fields and energy fluxes in the CGHS model}
\label{QFT-CGHS}

In this section, we outline the basic framework for studying the quantum real scalar field on the CGHS black hole and calculate the Hawking flux in asymptotic infinity,  and the negative flux through the horizon.

\subsection{Field quantization and Hawking radiation}

The standard framework of Quantum Field Theory (QFT) in curved spacetime is  simplified   when there  are  asymptotic regions where one can expand the field in  some  appropriate  canonical basis of  mode functions. In this case we will consider ${\mathscr I}^{-}_{L}$ and ${\mathscr I}^{-}_{R}$ as our asymptotic {\it in} region, and the  interior of the black hole plus  the  ${\mathscr I}_{R}^+$ region as our asymptotic {\it out} region (although   in the interior black hole region there is no time-like Killing field, and thus no  natural notion of particle  and  no canonical modes in terms of which to perform the quantization).
We are interested in expanding the field $f$ in these two regions of   the spacetime. This will allow  one  to find the energy fluxes in the following subsection.

In the {\it in} region the field operator can be expanded as
\begin{equation}
\hat{f}(x)=\sum_{\omega}(\hat{a}_{\omega}^{R} u_{\omega}^{R} + \hat{a}^{R \dag}_{\omega} u_{\omega}^{R*} + \hat{{a}}^{L}_{\omega} u_{\omega}^{L}+\hat{{a}}^{ L \dag}_{\omega} u_{\omega}^{L*}),
\label{foin}
\end{equation}
where, the basis of functions (modes) are as follows:
\begin{equation}
u_{\omega}^{R}=\frac{1}{\sqrt{4\pi\omega}}e^{-i\omega y^{-}}
\label{ur}
\end{equation}
and
\begin{equation}
u_{\omega}^{L}=\frac{1}{\sqrt{4\pi\omega}}e^{-i\omega y^{+}},
\label{ul}
\end{equation}
with $\omega>0$. The superscripts $R$ and $L$ refer to the  right and left moving modes. These modes will define an \emph{in} vacuum right ($\kets{0}^{in}_{R}$) and \emph{in} vacuum left ($\kets{0}^{in}_{L}$) whose tensor product ($\kets{0}^{in}_{R}\otimes \kets{0}^{in}_{L}$) will define our {\it in} vacuum.

We can also expand the field in the {\it out} region in a manner similar to eq.  (\ref{foin}). In this region, the complete set of modes  include those   that have support on the  outside {(region II in Fig. \ref{cghs}) and on the  inside (region III in Fig. \ref{cghs})} the event horizon.  Therefore, the field operator has the following form
\begin{eqnarray}
\hat{f}(x) &=& \sum_{\omega}(\hat{b}_{\omega}^{R} v_{\omega}^{R} + \hat{b}^{R \dag}_{\omega} v_{\omega}^{R*} + \hat{{b}}^{L}_{\omega} v_{\omega}^{L}+\hat{{b}}^{ L \dag}_{\omega} v_{\omega}^{L*}) + \notag \\
&& \sum_{\tilde \omega}(\hat{\tilde{b}}_{\tilde \omega}^{R} \tilde{v}_{\tilde \omega}^{R} + \hat{\tilde{b}}^{R \dag}_{\tilde \omega} \tilde{v}_{\tilde \omega}^{R*} + \hat{\tilde{b}}^{L}_{\omega} \tilde{v}_{\tilde \omega}^{L}+\hat{\tilde{b}}^{ L \dag}_{\tilde \omega} \tilde{v}_{\tilde \omega}^{L*}).
\label{fout}
\end{eqnarray}
Hereafter, the modes and operators with and without tildes   correspond to those  associated  with the regions  inside and outside the horizon, respectively. Note that there is an arbitrariness in the choice of basis inside the horizon, as there is no  timelike Killing field,  and   thus  no  canonical definition of particles there. However, this arbitrariness is  not  expected to affect  the physical results  we  will  be  interested on, as  the  specific  states inside the black hole  should not  be relevant  to   any  of  the quantities of interest,  which  will    be   related  to  things  that are, in principle,   observable  by asymptotic   observers\footnote{The precise  study of this  question  will be left  for  future work.}. The convenient basis of modes in the exterior  (region-II in  Fig. \ref{cghs}) are the following:
\begin{equation}
v_{\omega}^{R}=\frac{1}{\sqrt{4\pi\omega}}e^{-i\omega\sigma^{-}}\Theta(-(x^{-} + \Delta))
\label{eq.5}
\end{equation}
and
\begin{equation}
v_{\omega}^{L}=\frac{1}{\sqrt{4\pi\omega}}e^{-i\omega\sigma^{+}}\Theta(x^{+} - x_0^{+}).
\label{eq.6}
\end{equation}
Similarly, following \cite{aFabjNs05, sbGidwmNel92}, one can define a set of   black hole  interior modes  {(region III in Fig. \ref{cghs})}.
For that,  we make use of $y^{\pm}$ coordinates which are well defined in region III. The left moving modes  (moving from region II to region III) are   simply a  continuation of each other in the two  regions,  because  they never cross the collapsing matter shell. However, the right moving modes coming from region I' to region III do cross the matter  shell,  and therefore will   generally  lead to non-trivial Bogolubov coefficients. The modes in region I' are  given by eq.  (\ref{ur}), whereas, in region III they can be chosen as \cite{sbGidwmNel92}:
\begin{equation}
{\tilde{v}}_{\omega}^{R} (y^-) \equiv {v}_{\omega}^{R*}(-y^-).
\label{vint}
\end{equation}
The above formula,  defining  the modes in region III,  involves  using  the expression  for $\sigma^- ( y^-) $ (from eq.  (\ref{smy}))and the substitution in the argument $y^-$ by $-y^-$.
Usually, the consideration of the  operator expansions eq.  (\ref{foin}) and eq.  (\ref{fout}), leads to  two sets of Bogolubov coefficients for the right moving sector. The first  corresponding  to the relations   between  the modes in regions I' and III,  and the second  to that between  the modes  in  regions I and  region  II. One   focuses  on the transformation from the  {\it in} to  the {\it exterior} modes ({\it i.e.} regions I and II) because  that   is   what leads to the Hawking flux.

In this  way  one obtains    the Hawking radiation in the asymptotic limit  (given by the right moving sector), and  the  negative flux at the horizon (given by the left moving sector).

In order to provide  an  appropriate  notion of late time particle production (in terms of normalizable modes), it is convenient  to replace the above delocalized plane wave type modes eq.  (\ref{eq.5}) and eq.  (\ref{eq.6}) by a complete orthonormal set of discrete wave packets modes, such as
\begin{equation}
v_{jn}^{L/R}=\frac{1}{\sqrt{\epsilon}}\int_{j\epsilon}^{(j+1)\epsilon}d\omega e^{2\pi i\omega n/\epsilon}v_{\omega}^{L/R},
\label{ulr}
\end{equation}
where the integers $j\ge 0$ and  $-\infty <n<\infty$. These  modes  correspond   to wave packets   which are peaked about $\sigma^{+/-}= 2\pi n/\epsilon$   and    which have a  width  $2\pi/\epsilon$.  Taking a  small value of epsilon ensures that the modes' frequency is narrowly centered around $\omega\simeq \omega_{j}=j\epsilon$. This, in turn, gives a clear physical interpretation of the count of a particle detector sensitive only to frequencies within $\epsilon$ of $\omega_{j}$, while switched on for a time interval $2\pi/\epsilon$ at time $2\pi n/\epsilon$. A  similar  procedure  is applied to  convert the modes eq.  (\ref{ur}) and eq.  (\ref{ul})  into  localized  modes  making up a  discrete basis. Writing the modes in discrete basis gives a natural definition of the field operators eq.  (\ref{foin}) and eq.  (\ref{fout}) in terms of an orthonormalized set up.

With these basis of modes in the \emph{in} and \emph{out} regions defined,  we can construct  the corresponding Fock space quantization of the field in  each  region. Using standard procedures \cite{Wald}, one can construct, for example, the Fock space for the right moving sector of the field in the \emph{exterior} region, $\mathscr{F}_{ext}^R$.  The Fock spaces for the \emph{in} and \emph{out} quantizations are, respectively,
\beq
\mathscr{F}_{in}= \mathscr{F}_{in}^R \otimes \mathscr{F}_{in}^L
\eeq
and
\beq
\mathscr{F}_{out} = \mathscr{F}_{int}^R\otimes \mathscr{F}_{int}^L\otimes \mathscr{F}_{ext}^R\otimes \mathscr{F}_{ext}^L \, .
\label{Fout}
\eeq
Now, consider the distribution of occupation numbers $F=\{\dots, F_{nj}, \dots\}$, $F_{nj}\geq0$ integer, such that $\sum_{nj} F_{nj}<\infty$ and the normalized state
\beq
\kets{F}_R^{ext} = C_F \prod _{nj}  (\hat{b}_{nj}^\dagger)^{F_{nj}} \kets{0}_R^{ext},
\eeq
where $C_{F}$ is a normalization factor. The set of all possible states of this form, $\{\kets{F}^R_{ext}\}$, constitutes a basis of $\mathscr{F}_{ext}^R$. Basis for all other Fock spaces can be constructed similarly.

Following  \cite{sbGidwmNel92},  one   writes  the {\it in} vacuum  state   as  a superposition of all the particle states of {\it out} basis. As we have noted, the non trivial Bogolubov coefficients occur only for the right moving modes and thus, one can expand formally $\kets{0}^{in}_R$ in the basis of the \emph{out} (\emph {exterior} and \emph{interior}) right moving sector's Fock space. This is the standard derivation of the Hawking radiation in the CGHS model, which  gives  \cite{sbGidwmNel92}
\footnote{ The thermal coefficients that appear in the form of the density matrix operator $\rho^{in}_R$ come from the explicit form of the Bogolubov coefficients evaluated at late time limit, which fits our purposes in this work. However, for the CGHS model they can be determined without taking this limit \cite{aFabjNs05}.}:
\beq
\kets{0}^{in}_R = N \sum_F e^{-\frac{\pi}{\Lambda} E_F} \kets{F}^{int}_{R}\otimes\kets{F}^{ext}_{R}\,,
\label{0inR}
\eeq
where  $N$ is a normalization factor and $E_F \equiv \sum_{nj} \omega_{nj} F_{nj}$ is the energy of state  $\kets{F}^{ext}_R$ with respect to late-time observers near $\mathscr{I}^+_R$ and $\sum_F \equiv \sum_{F_{nj}}\sum_{F_{n'j'}}\dots$ where all the sums run from $0$ to $\infty$.
Then, the full \emph{in} vacuum can be expressed as
\begin{eqnarray}
\kets{0}^{in} & =& \kets{0}^{in}_{R}\otimes \kets{0}^{in}_{L} \notag\\
                      &=& N \sum_F e^{-\frac{\pi}{\Lambda} E_F} \kets{F}^{int}_{R}\otimes\kets{F}^{ext}_{R} \otimes \kets{0}^{in}_{L} \,.
\label{invac}
\end{eqnarray}
In the above expression we have used the fact that the vacuum state for the left movers is unchanged in both quantizations (due to trivial Bogolubov transformations).

With that, we are in a position to move to the remaining part of our work using the above CGHS model as our playground. We start by finding the energy fluxes by calculating the renormalized
energy momentum tensor in the {\it in} vacuum.


\subsection{Energy fluxes}\label{sec:enfl}
Here we are interested in finding the renormalized energy momentum tensor in the {\it in} vacuum. For that we follow the method introduced by Davies, Fulling and Unruh (DFU) \cite{dfu}, and also used by Hiscock \cite{waHis81}\footnote{For a extensive discussion of quantum fields on two dimensional black holes see, for instance, \cite{aFabjNs05}.}.

This method uses the property that in two dimensions any spacetime metric can be written in conformally flat coordinates as
\begin{equation}
ds^2 = C(u, v) du dv
\label{conf}
\end{equation}
where $u$ and $v$ are null coordinates. One can always introduce another set of null coordinates $\overline{u}$ and $\overline{v}$ such that $\overline{u} = \overline{u} (u)$ and $\overline{v} = \overline{v} (v)$. In the DFU method, one defines a vacuum state   directly in  terms of   these $(\overline{u}, \overline{v})$ coordinates. In the {\it in} region, this vacuum corresponds to  the $in$ vacuum (defined before the gravitational collapse), since in this region the  metric is flat,  and one has $\overline{u} = u, ~ \overline{v} = v, ~ C(\overline{u},~\overline{v}) =1$. However, in the {\it out} (defined long after the gravitational collapse) region this is no longer the vacuum, since $\overline{u}$ and $\overline{v}$ are nontrivial functions of $u$ and $v$. As a consequence, one finds particle creation and non-zero energy fluxes with respect to the {\it in} vacuum. The renormalized energy-momentum tensor in the {\it in} vacuum is given by \cite{dfu}
\begin{eqnarray}
\langle 0|^{in} T_{\mu\nu} | 0 \rangle^{in} = \theta_{\mu\nu} + \frac{R}{48\pi} g_{\mu\nu}
\label{emtens}
\end{eqnarray}
with
\begin{eqnarray}
\theta_{\overline{u}~\overline{u}} &=& -\frac{1}{12\pi} C^{1/2} \partial_{\overline{u}}^2(C^{-1/2}) \\
\theta_{\overline{v}~\overline{v}} &=& -\frac{1}{12\pi} C^{1/2} \partial_{\overline{v}}^2(C^{-1/2}) \\
\theta_{\overline{u}~\overline{v}} &=&  \theta_{\overline{v}~\overline{u}} = 0\\
C &=& C(\overline{u},\overline{v}).
\end{eqnarray}

Using  the above expressions one can  obtain the  explicit expressions   for  the renormalized energy-momentum tensor for CGHS model. First  for  the {\it in} linear dilaton vacuum region, we  have the  metric eq.  (\ref{iny}), (region I and I' in Fig. \ref{cghs}) and  thus one   has   null coordinates  readily available $\overline{u} = y^{-}$ and $\overline{v}=y^{+}$ and $C(y^{-}, y^{+}) =1$. Whereas, for the {\it out} region (region II and III in Fig. \ref{cghs})  the metric  is  given by eq.  (\ref{outy}), and  thus  the conformal factor  is:
\begin{equation}
C(y^{-}, y^{+})= -\frac{1}{e^{\Lambda y^- \left(\frac{M}{\Lambda^2 \Delta}e^{-\Lambda y^+} + e^{-\Lambda y^-} -1\right)}}.
\end{equation}
The resulting  components of energy-momentum tensor in $y^{+}, ~ y^{-}$ coordinates are thus,
\begin{eqnarray}
\langle 0|^{in} T_{{y^{+}}y^+} | 0 \rangle^{in} &=& -\frac{\Lambda ^2 M e^{\Lambda  {y^{-}}} \left(M e^{\Lambda  {y^{-}}}+2 \Delta  \Lambda ^2 e^{\Lambda  {y^{+}}}-2 \Delta  \Lambda ^2 e^{\Lambda  ({y^{+}}+{y^{-}})}\right)}{48 \pi  \left(M e^{\Lambda  {y^{-}}}+\Delta  \Lambda ^2 e^{\Lambda  {y^{+}}}-\Delta  \Lambda ^2 e^{\Lambda  ({y^{+}}+{y^{-}})}\right)^2}, \\
\langle 0|^{in} T_{y^-y^-} | 0 \rangle^{in} &=& \frac{\Lambda ^2 e^{\Lambda  {y^{-}}} \left(\Delta  \Lambda ^2 e^{\Lambda  {y^{+}}}-M\right) \left(M e^{\Lambda  {y^{-}}}+2 \Delta  \Lambda ^2 e^{\Lambda  {y^{+}}}-\Delta  \Lambda ^2 e^{\Lambda  ({y^{+}}+{y^{-}})}\right)}{48 \pi  \left(M e^{\Lambda  {y^{-}}}+\Delta  \Lambda ^2 e^{\Lambda  {y^{+}}}-\Delta  \Lambda ^2 e^{\Lambda  ({y^{+}}+{y^{-}})}\right)^2}. \nonumber\\
\end{eqnarray}

Next  we  write  the relevant  expression  for the  energy fluxes  in region II of Fig. \ref{cghs}  in terms of  the  $\sigma^{+},~\sigma^{-}$ coordinates. These  correspond to   $\langle 0|^{in}  T_{\sigma^{\pm} \sigma^{\pm}} | 0 \rangle^{in} = (\frac{\partial y^{\pm}}{\partial \sigma^{\pm}})^2 \langle 0|^{in}  T_{{y^{\pm}}y^{\pm}}  | 0 \rangle^{in}$, thus  giving,
\begin{eqnarray}
\langle 0|^{in} T_{\sigma^{-} \sigma^{-}} | 0 \rangle^{in} &=& \frac{\Lambda ^2 e^{\Lambda  {\sigma^{-}}} \left(\Delta  \Lambda ^2 e^{\Lambda  {\sigma^{+}}}-M\right) \left(M e^{\Lambda  {\sigma^{-}}}+\Delta  \Lambda ^2 e^{\Lambda  ({\sigma^{-}}+{\sigma^{+}})}+2 \Lambda  e^{\Lambda  {\sigma^{+}}}\right)}{48 \pi  \left(\Delta  \Lambda  e^{\Lambda  {\sigma^{-}}}+1\right)^2 \left(M e^{\Lambda  {\sigma^{-}}}+\Lambda  e^{\Lambda  {\sigma^{+}}}\right)^2}, \nonumber\\
\langle 0|^{in} T_{\sigma^{+} \sigma^{+}} | 0 \rangle^{in} &=& -\frac{M \Lambda^2 e^{\Lambda\sigma^{-}} (M e^{\Lambda\sigma^{-}} + 2\Lambda e^{\Lambda\sigma^{+}})}{48\pi (M e^{\Lambda\sigma^{-}} + \Lambda e^{\Lambda\sigma^{+}})^2}.
\end{eqnarray}
In the asymptotic limit ($\sigma^{+} \rightarrow \infty$), the flux at ${\mathscr I}_R^{+}$ simplifies to
\begin{eqnarray}
\langle 0|^{in} T_{\sigma^{-} \sigma^{-}}| 0 \rangle^{in} = \frac{\Lambda^2}{48\pi} \left(1-\frac{1}{(1 + \Delta\Lambda e^{\Lambda\sigma^{-}})^2}\right),
\end{eqnarray}
corresponding  to the results  in \cite{CGHS92,sGidd92}. In the late time limit {(see Fig. \ref{cghs})}, this gives the Hawking flux
\begin{eqnarray}
\langle 0|^{in} T_{\sigma^{-} \sigma^{-}}^{H} | 0 \rangle^{in} = \frac{\Lambda^2}{48\pi}.
\end{eqnarray}
On the other hand,  near the horizon (in the limit $\sigma^{-} \rightarrow \infty$), one finds
\begin{eqnarray}
\langle 0|^{in}T_{\sigma^{+} \sigma^{+}}| 0 \rangle^{in} = - \frac{\Lambda^2}{48\pi}
\end{eqnarray}
\emph{i.e.}, an equal amount of negative flux going inside the black hole. As a result  of these fluxes  the  black hole loses energy during  its evaporation.

These basic features  of the model   will be important in our discussion of the  end  state  of the black hole   evaporation  process.


\subsection{Comment on renormalization and Hadamard form}
\label{Haddamard}

In the standard approach to QFT in flat spacetime, we have an entirely satisfactory prescription of renormalization of the energy-momentum tensor on a suitable class of states in the standard Fock representation. This is given in terms of ``normal ordering", which is  well defined  due to the fact that  there is a unique  canonical  vacuum state  connected to the  notions of  energy used  by all inertial observers. The key problem in extending  this idea to curved spacetime is due to the absence of a ``preferred'' vacuum state. Moreover, even if one chooses such ``preferred states", for example, in the case of stationary spacetimes, there is always ``vacuum polarization" that makes $\langle T_{\mu\nu} \rangle \ne 0$. As a result, normal ordering is not a good prescription for renormalization in curved spacetime.  Thus, one  needs  a more general prescription that can be extended to curved spacetime. Fortunately, such an extension exists and it is given by the ``Hadamard renormalization".

The essence of the  Hadamard renormalization \cite{lPardTom07} for the real scalar field $\phi(x)$ is to find out the physically relevant states $ \{ |\psi\rangle \}$
 in the standard Fock space such that the difference $F(x,x') \equiv \frac{1}{2}(\langle \psi | \phi (x) \phi (x') |\psi \rangle + \langle \psi | \phi (x') \phi (x) |\psi \rangle) - H (x,x')$ is a smooth function of $x$ and $x'$. Here $H(x,x')$ is the ``Hadamard ansatz'' for the Green's function whose precise form, even for the real scalar field, varies depending on the dimensionality. By subtracting this term, one removes  the singular behavior in $G^{(1)} (x,x') = \frac{1}{2}(\langle \psi| \phi (x) \phi (x') |\psi \rangle + \langle \psi| \phi (x') \phi (x) |\psi \rangle)$ {\it if} and {\it only if} $| \psi \rangle$ is a Hadamard state. In other words, if the singular structure of the two point function is purely ``Hadamard", one obtains a well defined {\it renormalized} two-point function. Moreover, it eventually gives a physically acceptable renormalized energy-momentum tensor $\langle \psi| T_{\mu\nu} (x) |\psi \rangle$ by: (i) taking appropriate derivatives of $F(x,x')$ with respect to $x$ and $x'$, (ii) taking the coincidence limit $x\rightarrow x'$, and (iii) making this compatible with Wald's axioms \cite{Wald} by adding or subtracting local curvature counter-terms.

The reason for using the {\it in} vacuum to calculate the fluxes is that this state is known to be a Hadamard state. Thus  the DFU approach \cite{dfu} should  be compatible with the Hadamard approach.
Thus,  one expects that  the  calculated Hawking fluxes  at infinity,  and  at the horizon  would  be  the  correct ones. (On the other hand, as we show in Appendix \ref{Hadamard} by a direct calculation,  the generic states  in the {\it int}, {\it out} or $ int\otimes out$ bases are {\it non-Hadamard}. These states are analogous to the  Boulware state, which is known to be divergent at the horizon. This divergence is  not of the Hadamard form. This is to say that, although the left hand side of eq.  (\ref{invac}) is Hadamard, the  individual terms in the sum on the right hand side are non-Hadamard.

Here,  we  might  become  very concerned    because,   when  the    state  of a  quantum field  is not of the  Hadamard  form,  one  can not    define,  in a reasonable  way,  a smooth
  renormalized    energy  momentum tensor   expectation value for it. Thus,   it   seems,    that  allowing  such  states to appear in the characterization of  the  evolution of our  fields, in  the black hole    space-time, as  we  will be  doing  in the following sections,   would    be   equivalent to  allowing    the kind  of dramatic  departures from   the smooth physics   that have been  characterized  as ``firewalls", in regions  where   no  drastic departures  from  semiclassical  gravitation   should  be expected.

     In fact, one can find   a  very   similar situation
  arising  in  even  more mundane situations: consider the Minkowski  vacuum  as  described in terms of   the (two  wedges) Rindler coordinates.  As it is  well known, this  state, when tracing
  over,  say,  the right wedge's  degrees of freedom,  corresponds to a thermal state.   Let us consider now  a detector  providing  the  measurement of the  number  of   particles  in a certain mode of the
  field.    The point is that  a   state   with  a  definite  number of  particles
  in the corresponding mode  (both in the left  and right  Rindler wedges)   is  not   a Hadamard  state, and thus   has  an ill  defined    expectation  value  for  the energy-momentum  (precisely at the  Rindler  horizon),  which  would have to be  considered  as a  singular state  there.
   Thus, if,  as  a result of the  measurement, the  state of the system becomes  one  such a state,  we  would
    have something similar to the  emergence of a firewall  at the Rindler  horizon\footnote{We thank  R.  Wald  for a very informative  discussion  regarding this  issue.}.  The conclusion   we  must draw  from  such analysis is   that no  detection,   that  could  be  modeled  in terms of a  smooth   interaction  between a  localized   detector  and the  quantum  field,  could  ever provide  a precise  measurement of the number of Rindler particles in any  given mode.

 The  above   discussion illustrates the   lesson  we  should  draw  regarding the situation   involving the  quantum   field  in  the   case of  a  black hole  subject to   the evolution characterized  by the modified  Schr\"odinger equation associated  with a  dynamical collapse theory:  If the   modification of   the evolution  equation   under consideration, involves   only smooth  local  operators,  it   would not, in  any finite  amount of time,  result in the collapse  of   the  state of the field  into a state with definite  number of particles in  any  of the modes that  are associated  with  unphysical  divergences  at the horizon.   We  will   have  more to say on this  issue in the following sections.

\section{ A word about pure, mixed and thermal states}  \label{sec:thermal-states}

  In  quantum  mechanics  one is often led to consider not just  vectors (or  more precisely  rays) in the Hilbert space,   as characterizing the state  of a system, but often  more general objects  known  as  density matrices  are used  for  that task.   The  cases   where  that occurs  involve   situations  where one  considers   ensembles  of identical  systems,    situations  in  which one does not  know    the  precise    state of the system, or when one   considers a  subsystem of a larger system.

  In  the practical usages one  very  seldom  distinguishes   among the above situations,  a  fact that leads to  a tendency  to simply and  generically ignore  the differences.
   However,   we believe  that  when considering  issues of principle  it is  essential to make  the  appropriate   distinctions  if one is to  avoid   generating confusion.
    One situation where the distinction is  very  important   concerns the   analysis of the measurement problem which is  often addressed  using decoherence arguments.

    One    simple  and  very illustrative example  is  provided  by  a  simple   EPR  pair of  spin $1/2$  fermions   in   zero  angular momentum  state.
      If we consider the  particles   moving    along the $z$ axis,  we  can  describe the state  of the system using the   basis  of  spin  states   oriented  along say the $x$  axis $\lbrace
     \kets{+1/2, x} ,    \kets{-1/2, x}    \rbrace$  for the Hilbert space of  each particle.  The   state of the   two particle system is then $\kets{\Psi} =\frac{1}{\sqrt{2}}  (\kets{+1/2, x}^{(1)} \otimes \kets{-1/2, x}^{(2)}
     +  \kets{-1/2, x}^{(1)} \otimes \kets{+1/2, x}^{(2)}   )$.  If we  decide to focus on the particle  $1$, and   thus   characterize  the situation with the reduced
       density matrix   $\rho^{(1)}  \equiv Tr_{2} (\kets{\Psi} \bras{\Psi} )  = \frac{1}{2}  (\kets{+1/2, x}^{(1)}  \bras{+1/2, x}^{(1)}
     +  \kets{-1/2, x}^{(1)} \bras{-1/2, x}^{(1)}   )$,   we  might be inclined to consider that  the particle $1$ is  now   in a definite state:  either  $ \kets{+1/2, x}$ or $   \kets{-1/2, x} $,  with probabilities  $1/2$  for  each   alternative.    There are at least two things  that  clearly indicate that  such interpretation  is  not  correct.  First, the   simple fact that   had we started describing the system  using the   basis  of  spin  states   oriented  along say the $y$  axis $\lbrace     \kets{+1/2, y},    \kets{-1/2, y}    \rbrace$  for the Hilbert space of  each particle, we  would have ended  with the    expression $\rho^{1}  \equiv Tr_{2} (\kets{\Psi} \bras{\Psi} )  = \frac{1}{2}  (\kets{+1/2, y}^{(1)}  \bras{+1/2, y}^{(1)} +  \kets{-1/2, y}^{(1)} \bras{-1/2, y}^{(1)}   )$,  and   according to the above we  would be entitled to  assert that  the particle  $1$ is    in a definite state:  either  $ \kets{+1/2, y}$ or $   \kets{-1/2, y} $,  with probabilities  $1/2$  for  each   alternative, which is  in clear   contradiction with the previous   conclusion. The  second  is  the    existence    of  the strong   non-classical  correlations,     (which  have  been   experimentally demonstrated by the famous   experiments of Aspect  {\it et. al.} \cite{Aspect})  make it  clear that, such  interpretations are untenable.  Thus, we conclude that,     the fact that  we  use   the same  mathematical objects to describe   various  physical   situations,  requires us to be extra   careful   to  avoid  the confusion of one such situation  with the other,   otherwise  we  could  very  quickly be  driven  to erroneous  conclusions.

 When considering   any system  one  can  then wonder if  it should be described  by a  pure  or mixed  state.  When  we can identify  that the system as part of another larger system, and  we know   there  are  correlations between  it and other parts,  it is  clear  that   it should be  described  by  a mixed    density matrix.  But,  what  should    be  the description if  one can not identify  the larger system to which  our system is a part of?  Or if  we   can point  to such  a larger system but we also know   there  are no  correlations  with the (sub-)system of interests? Should  we   consider  that  the  appropriate  description must be pure,  and that,  if we   consider a density matrix  it  is only because of our ignorance about  that pure state?   The  problem  becomes  specially acute  when  thinking of the universe  as a whole,  a  situation in  which,  by definition, there can be  no larger system.  Is the  universe  necessarily in a pure state? Or can it be in a  mixed  state which reflects something other than our ignorance  about the  particular pure state the Universe is?

In  order to  make progress    we    will  adopt a   rather  conservative viewpoint  and   adhere  to it consistently throughout the     coming  discussions. That is, in this   work   we  will take the view  that   individual isolated    systems   that  are  not entangled  with others  systems  are represented  by  pure states, and   can  be represented    either  by  the corresponding density matrices   that satisfy $\rho^2 =\rho$,  or by the     unique  ray  $\kets{\psi} $ (the phase is  irrelevant)  for which  one can write  $\rho= \kets{\psi}\bras{\psi}$.

    \smallskip

   The  case  where  one    is  concerned  with an individual  system   and  one does not know  exactly what  state  the system is in,  is  in fact  a  particular   situation   of   the  usage of an  ensemble.
   In  such   situations,  as  in   most usages of probabilistic  considerations, the  ensemble is  employed to represent  our lack on knowledge regarding the system's state. That  is the  case   even if one  is dealing, in  practice,  with a  unique specific   system. This is just what is done,  for instance,   in  making probabilistic  considerations in   weather prediction.   So  regarding issues of principle  we have, in fact, to distinguish only  between two kinds  of situations  as  far  as the  characterization of a system   via  a general  density  matrix.
 Thus,   mixed states  will   occur  when  one   considers either:

          \smallskip

 { a)}   An  ensemble  of  (identical) systems  each in a pure state. These  are the ``proper" mixtures.

          \smallskip

{ b)}  The  state  of a subsystem  of a larger  system  (which is  in a pure  state),  after  we {\it ``trace over" } the rest of the system. These  are the ``improper" mixtures.
      \smallskip

      The   above    characterizations of   mixed  states  as    \emph{proper}  or \emph{improper} follows from the   terminology introduced by  d{}' Espagnat \cite{dEspagnat}.
     An {ordinary  (quantum) thermal state}, (such as what occurs  in statistical mechanics) represents  an  ensemble,  where the weights  are  simple  functions, characterizable  by
     temperature and chemical potentials (e.g.   the ensemble that  characterizes   a gas   at   room temperature) and is thus  a proper thermal state.
     An    {``improper"   thermal state}  is     an  improper mixed  state   where the weights  happen  to be  thermal (e.g.  the Minkowski  vacuum, described in Rindler coordinates,  after   tracing over the other wedge).

From the above point of view,  which  seems to be the most   demanding in the present   context,  resolving the
BH  information paradox  would  require explaining  how  a pure state becomes  a  ``proper''
 thermal state rather than a ``improper"  one,  simply  because the    region corresponding  to black hole interior   will disappear.

\section{Brief review of the CSL theory} \label{sec:CSL}

We  start  our discussion by  presenting a  particularly  simple form of CSL, which describes collapse towards one or another  of the eigenstates of an operator $\hat A$ with rate $\sim\lambda$. We leave  aside,   for the moment, the question regarding   what  dictates  the  selection of such operator\footnote{As  we  will  see later, one   envisions  a truly fundamental theory of   collapse   that  involves  a  general rule  determining such operator in  all situations,     in  terms of  the fundamental  physical degrees  of freedom,   \emph{i.e.}  the fields of the standard  model  and,  in order  to  be really universal,   in term of whatever  is the fundamental   quantum description of gravity at the quantum level.
}.   Again, the  work in this  section is  just a  review  of existing work, and  contains nothing original.  For a very pedagogical   and  detailed  review  we  suggest  turning to \cite{moreCSL, cslnew}.
Those  readers familiar with the model can safely  proceed  directly to the next section.


In using the  theory,   one   needs to   consider  two equations \cite{cslnew}.
The first is a stochastically modified  version  of Schr\"odinger equation,  which in the simple case  where  we  take   the  hamiltonian  $ H\equiv 0 $  has,  as a    general solution:
\beq \label{cslb}
|\psi, t\rangle = e^{-\frac{1}{4\lambda t} [B(t) - 2\lambda t \hat A]^2}|\psi, 0 \rangle
\eeq
  where   $B(t)$  is  a stochastic  function of time{\footnote{ $B(t)$ corresponds to the Wiener  process appearing  in the  corresponding Langevin equation in the  standard treatment of Brownian motion.}},  $\hat A $  is a  hermitian  operator and  $\lambda$   is a  positive   valued    parameter. The second is the probability rule   for   the specific  realization of    the function $B(t)$. That   rule  can be  presented  in terms of  the probability that
   the  specific  realization at   each  time  $t$  has the value within $B(t)$ to $B(t) + dB(t)$, given by
\beq \label{cslpb}
P[B(t)] dB(t) = \frac{dB(t)}{\sqrt{2\pi\lambda t}} \langle \psi,t | \psi, t\rangle,
\eeq
with the assumption that the initial state has a unit norm. These two equations define CSL and everything else is derived using these two.

\noindent When  including a  nontrivial   Hamiltonian,  the state vector dynamics  is most easily understood in terms of   small   individual time  ``steps". For that   we  consider an ``infinitesimal"  time interval  of duration  $dt$, and  state the evolution equation \eqref{cslb}    during that interval:
\beq\label{CSLnew1}
|\psi,t\rangle =  e^{-dt \big[i\hat H+\frac{1}{4\lambda}[w(t)-2\lambda\hat A]^{2}\big]}|\psi,t-dt\rangle,
\eeq
where $w(t) = dB(t)/dt $ is a random white noise function. The  probability that  its value   at $t$  lies  within the  interval  $(w(t), w(t) + dw(t))$ is now:
\beq \label{CSLnew2}
P(w) dw = \frac{\langle\psi,t|\psi,t\rangle}{\langle\psi,t-dt|\psi,t-dt\rangle} \frac{dw(t)}{\sqrt{ 2\pi\lambda/dt}}
\eeq
Over a finite time interval, say from $t=0$ to $t$ in the step of $dt$,  \eqref{CSLnew1} takes the following form
  \begin{equation}\label{CSL1}
|\psi,t\rangle={\cal T}e^{-\int_{0}^{t}dt'\big[i\hat H+\frac{1}{4\lambda}[w(t')-2\lambda\hat A]^{2}\big]}|\psi,0\rangle,
\end{equation}
(where ${\cal T}$ is the time-ordering operator) and the probability rule  again has to be  described  by a   $  N\to  \infty $ limiting process  involving dividing the  interval $(0,t)$  on   $N$    steps   $ (t_0= 0,  t_1,.....  t_N =t) $    with   duration  $ dt_i =  t_i-t_{i-1}= dt$.  The probability that   the  function  $w(t)$ lies  in the tube  $Dw(t)$  characterized  by the restriction that in the $i$-th step the white noise function takes a value between $w(t_i)$ and $w(t_i) + dw(t_i)$ is  given by:
\begin{equation}\label{CSL2}
P(w) Dw(t) = \langle\psi,t|\psi,t\rangle\prod_{i=1}^{N}\frac{dw(t_{i})}{\sqrt{ 2\pi\lambda/dt}}.
\end{equation}
\noindent  Thus now the probability rule \eqref{CSL2} is a joint probability distribution over the entire time interval $t=0$ to $t$. In deducing \eqref{CSL2}  from \eqref{CSLnew2}, one has to assume that the  norm of the  initial  state vector was unity. For  later times the  state vector norm evolves dynamically (\textit{not} equal 1), so eq. (\ref{CSL2}) says that the state vectors with largest norm are most probable.
 Note that  the interpretation of the theory  does not  require the  state vector to have norm  one  as  probabilities are now   assigned to  the realization of  the   stochastic  functions,    rather than to  the standard  quantum mechanical  amplitudes.  For further   discussion  see \cite{moreCSL, cslnew}.

 It is  now  straightforward to   see  that the total probability   for  realization of an arbitrary  stochastic  function is 1, that   is
\begin{eqnarray}\label{CSL3}
\int PDw(t)&=&\int Dw(t-dt)\int_{-\infty}^{\infty}\frac{dw(t)}{\sqrt{ 2\pi\lambda/dt}}\langle\psi,t-dt| e^{-dt'\big[-i\hat H+\frac{1}{4\lambda}[w(t')-2\lambda\hat A]^{2}\big]} \nonumber\\
  && e^{-\int_{0}^{t}dt'\big[i\hat H+\frac{1}{4\lambda}[w(t')-2\lambda\hat A]^{2}\big]}| \psi,t-dt\rangle\nonumber\\
&=&\int Dw(t-dt)\int_{-\infty}^{\infty}\frac{dw(t)}{\sqrt{ 2\pi\lambda/dt}}\langle\psi,t-dt| e^{-\frac{1}{2\lambda}[w(t')-2\lambda\hat A]^{2}} |\psi,t-dt\rangle\nonumber\\
&=&\int Dw(t-dt)\langle\psi,t-dt|\psi,t-dt\rangle= ... =\langle\psi,0|\psi,0\rangle=1.
\end{eqnarray}

 As  we indicated, the dynamics  is  designed to drive any initial state towards   one of  the  eigenstates $|a_{n}\rangle$  of $\hat A$.  We can see this  by considering the   simplified  case where   $\hat H=0$.   As  usual,   we  can   write   the initial  state in terms of those  eigenstates as
  \begin{equation}\label{CSL4a}
   |\psi,0\rangle=\sum_{n=1}^{N}c_{n}|a_{n}\rangle
   \end{equation}
  and thus,
 according to eqs. (\ref{cslb}), (\ref{cslpb}),  we  find
  \begin{eqnarray} \label{CSL4}
   |\psi,t\rangle &=& \sum_{n=1}^{N}c_{n}|a_{n}\rangle e^{-\frac{1}{4\lambda t}[B(t)-2 \lambda ta_{n}]^{2}},\\
 P[B(t)] dB(t)&=& \frac{dB(t)}{\sqrt{2\pi\lambda t}} \sum_{n=1}^{N}|c_{n}|^{2}e^{-\frac{1}{2\lambda t}[B(t)-2\lambda t a_{n}]^{2}},\label{CSL5}
\end{eqnarray}

\noindent According to eq. (\ref{CSL5}), the probability is then a sum of gaussians, each drifting by an amount $\sim a_{n}t$, and having a width $ \sim\sqrt{\lambda t}$.  Therefore,
after a while, the  result  can be  described  as a sum of  essentially separated gaussians.  Then, we  can identify   the  various  ranges  of  values of  $B(t)$  as   corresponding  to each  one of the possible outcomes.   If
$-K\sqrt{\lambda t}\leq B(t)-2\lambda ta_{n}\leq K\sqrt{\lambda t}$,  ($K>1$ is some suitably large number), the associated probability integrated over this range of  $B(t)$
is essentially $|c_{n}|^{2}$, and the state vector given by eq. (\ref{CSL4}) becomes  $|\psi,t\rangle\sim |a_{n}\rangle$.

Note that, when $\hat H\neq 0$, the hamiltonian dynamics interferes with the collapse dynamics, and  sometimes  the  full collapse is  never achieved, and  all that
happens is  a relative  narrowing  of the wave function about the   eigenstates of $\hat A $  leading to   a   kind   of  equilibrium  stage  between the two competing dynamics:  the   specific CSL  dynamics that  tends to  sharpen  the  wave  function  about one of the  eigenstates,   and the characteristic  Schr\"odinger  behavior  associated  with the spreading of the  wave function (such as what   occurs  with the position).   This  is    mathematically analogous to  what  one  expects when considering in classical physics  a cloud of   gas  subject both to the random fluctuations  and   the  effects of   gravity. One  part of the dynamics   tends to spread the  gas  cloud and the  other  tends to contract     it. Note however  that  in the above example the roles of the deterministic and  stochastic  components of   the dynamics, regarding respectively  contraction and diffusion (or  spatial  spreading), are interchanged  in   comparison   with the case of  a single  free particle  and a  CSL    modified dynamics with   a  smeared position operator driving the collapse (\emph{i.e.}  playing the role of the operator    $\hat A$ in  eq.  \ref {CSL1}).


The above, describes in  full the  evolution of an individual system ``prepared"  in  the initial state: the   presence of  the stochastic  function $w(t)$,   however,   clearly   renders the evolution highly  unpredictable,  and  all one can say  is that  the system  will be driven towards  one of the     eigenstates of $\hat A$. It is  often useful to discuss the fate of  an  ensemble of identically prepared  systems.  In that case, 
 we  consider a  collection of systems all prepared in the same initial state and  evolve  each one  of  them  according to  CSL  dynamics,
 which   tries to collapse  the  state vectors towards eigenstates of $\hat A$.  The result  is  an ensemble of differently evolved   state vectors, each characterized by a different  $w(t)$.
It is useful to have an expression for the density matrix which describes the ensemble of evolutions.  We obtain  that  simply by    considering  the  density  matrix  describing the   ensemble of  vectors  evolved  using eq. (\ref{CSL1}):
\begin{eqnarray}
\rho(t)&=&\int PDw(t)\frac{|\psi,t\rangle\langle\psi,t|}{\langle\psi,t|\psi,t\rangle}=\int Dw(t)|\psi,t\rangle\langle\psi,t|\nonumber\\
&=&\int Dw(t){\cal T}e^{-\int_{0}^{t}dt'\big[i\hat H+\frac{1}{4\lambda}[w(t')-2\lambda\hat A]^{2}}\big]|\psi,0\rangle\langle\psi,0|
e^{-\int_{0}^{t}dt'\big[-i\hat H+\frac{1}{4\lambda}[w(t')-2\lambda\hat A]^{2}\big]}\nonumber\\
&=&{\cal T}e^{-\int_{0}^{t}dt'\big[i(\underrightarrow{\hat H}-\underleftarrow{\hat H}]+\frac{\lambda}{2}[\underrightarrow{\hat A}-\underleftarrow{\hat A}]^{2}\big]}\rho(0),
\label{CSL6}
\end{eqnarray}
\noindent where the arrows $\rightarrow$, $\leftarrow{}$ under the operator indicate that the operator acts, respectively, on the right or on the left of $\rho(0)$, and the ${\cal T}$  reverse-time-order operator to the right of $\rho(0)$. It can be  shown  that the resulting evolution equation for the density matrix is
\begin{equation}
\frac{d}{dt}\rho(t)=-i[\hat H,\rho(t)]-\frac{\lambda}{2}[\hat A,[\hat A,\rho(t)]].
\label{CSL7}
\end{equation}
\noindent Therefore,  the ensemble expectation value of any operator$\overline{\langle\hat O\rangle}=\hbox{Tr}\hat O\rho(t)$  evolves  in time  according to
\begin{equation}
\frac{d}{dt}\overline{\langle\hat O\rangle}=-i\overline{[\hat O,\hat H]}-\frac{\lambda}{2}\overline{[\hat A,[\hat A,\hat O]]}.
\label{CSL8}
\end{equation}

In this  case,  one can  see  that the   Born rule is  recovered  in the sense that the portion of the  systems   that ends  up in the eigenstate turns
out  to be precisely the quantity obtained by projecting the initial state  on the   corresponding  eigenstates (the theory is  designed to  do
this).

     Of course,   one  could    now  ask: if we   still have to  determine  what   is  a measurement, and  decide,  in  each situation, what  is   the relevant operator $\hat A$,   and  determine a phenomenologically    suitable value   of  $\lambda$, {\it what has been  gained?}
The  beauty of the  proposal  is,   however, that it intends  to  use  the  above  mathematical   framework in connection with a
   a general     specification  of  the collapse operator that  would cover all situations. The existing proposals, framed in the  language of many particle  quantum mechanics,  the  operator is the
   tensor   product of  suitable  smeared position operators  for  all   the particles. What seems to make   this  possible  is the observation  that all measurement
   situations   can be  seen as  essentially  position measurements of   a    macroscopically  large  aggregate of particles  making, say,  a pointer  (or   in modern devices  something like  the
   corresponding position arrangement of   a large  number of   electrically charged  particles). The idea  is  then, that a measurement involves   in an   essential way the entanglement of the
   subsystem under  observation  with the  position coordinates of these large number of particles that constitute the pointer and     as  any of those  particles undergoes  collapse in their position,
   the  complete  system  collapses into a  state   for  which the subsystem under observation is in an eigenstate  of  the  measured quantity.
  This  of course can only  occur as long as the measurement interaction is  strong enough, \emph{i.e.} it leads to a strong  entanglement of the   quantity  being   measured    with the   fundamental position of  the  physical  system that  constitutes  the measuring  device.

 Thus,  the  CSL  proposal  involves the  selection of a  properly smeared  position operator for each particle  and  a  universal   value of the   parameter $\lambda$.
 In  fact,    later  refinements  indicate  that    the value of $\lambda$  should  be  universal   except for the fact that   each  type of particle  should  have  its own   value which, the    detail analysis suggests,  should be roughly proportional to the    square  of the particle's mass,   \emph{i.e.}  $ \lambda _i  =  \lambda_0 (m_i /  m_N)^2$   where        $  \lambda _i $  is   the collapse  rate for the smeared position of the particles of  the
 $ i^{th}$ species, $\lambda_0$  is  a universal parameter,   $m_i$   is the mass  of the
 $ i^{th}$ species of particles   and   $  m_N$  is  a fiduciary mass  scale   usually taken as  the  nucleon  mass ($1 GeV$).

         Of course, this can not be   considered  yet a satisfactory   candidate  for a  fundamental theory.   After all, we know that particles are not  the fundamental  entities. In fact,  Quantum Field  Theory   teaches  us  that    the fundamental  objects are   the fields,  and  that  the  notion of
  particle   is,  in general ,ambiguous  and  tied  intrinsically with the  particular symmetries of the  space-times  and   the coordinates we  use to describe them  (see for instance, \cite{lPardTom07, Wald}). Thus,   any candidate  for a truly fundamental  replacement of quantum theory will  likely  need to  be formulated in terms of quantum fields. It is  worth  mentioning  here,
  anticipating   latter  discussions, that   there do  exist  some  such  proposals,   involving   relativistic versions of  dynamical collapse theories.

\section{Application of CSL to the  CGHS model} \label{sec:appl_CSL}

In this section, we   describe the  CSL  evolution of states of  the real scalar  quantum  field living on the CGHS black hole.  The basic  idea  is  to  work in an interaction picture where  the  free  evolution  is  encoded  in the  quantum field operators and the CSL  effects are treated  as  an interaction and  codified in the  evolution of the quantum states.

 The results of this section will be used to support our main statement  in the next section.

\subsection{The foliation of CGHS spacetime}
\label{subsec:foliation}
In order to consider  the  CSL  evolution   of states of our quantum field  on the CGHS background using the   interaction picture  mentioned  above, we have to foliate the space-time,
 in  order to  have a well defined evolution operator  connecting the  states  associated  with   the  different specific  hypersurfaces. In this subsection,  we describe  such a  foliation.

As  described  in section \ref{sec:CGHS}, the space-time metric in the null Kruskal coordinates ($x^+,~x^{-}$)
is given in (eq. \ref{eq:1}). Now we  introduce the Kruskal coordinates $$T = \frac{x^+ + x^- + \Delta}{2}, \quad X = \frac{x^+ - x^- - \Delta}{2}.$$  The  metric   in the region  involving the black hole,  both its  exterior   and interior  (regions  II   and III in Fig. \ref{cghs}  respectively)  can be  written as
\begin{equation}
ds^2=-\frac{dT^2 - dX^2}{\frac{M}{\Lambda}-\Lambda^2 (T^2-X^2)}, (-\infty \le T \le \infty, -\infty \le X \le \infty ).
\label{metTX}
\end{equation}
These  coordinates  can  be   related with Schwarzschild-like time $t$ and space $r$ coordinates in the following way
\begin{eqnarray}
\tanh(\Lambda t) = \frac{T}{X},
\label{tout}\\
-\frac{1}{\Lambda^2} (e^{2\Lambda r} - M/\Lambda) = T^2 - X^2,
\label{rin}
\end{eqnarray}
where the time coordinate $t$ is well defined only in region II similar to the Schwarzschild case. Using eq. (\ref{ricci}) and the coordinate transformations for $T$ and $X$ as defined above, one finds $R= \frac{4M\Lambda}{M/\Lambda -\Lambda^2(T^2 - X^2) }$. Therefore,  the singularity is located  at $M/\Lambda = \Lambda^2(T^2 - X^2) $.

 Now,   we proceed to define  our foliation  $\Sigma_\tau $ and  related  coordinates $(\tau,\zeta)$  covering  the  regions  II  and  III. The idea is to   define   the hypersurface  $\Sigma_\tau $  by following in the region III ($e^{2\Lambda r} < M/\Lambda$) a  curve with  $ r=const.$, given by eq.  (\ref{rin}) and in the region II a $t=const.$ line, given by eq.  (\ref{tout}), connecting them via a line   $T =const.$.

  The  prescription is    defined once    we  provide the    joining  conditions  for   the   above   recipe. That  will be specified by two functions   $ T_1(X)$  and  $T_2(X)$  determining the   points  where the matching  takes place. The images of $ T_1(X)$  and  $T_2(X)$   will be  located in regions III and  II respectively.

More   specifically   the construction is  as follows:

As a  first  step  we   choose  any value of  $\tau$  in the   range  $ (0, \tau_s = \sqrt{M/\Lambda^3})$.  Then   we  start the    $\Sigma_\tau $ following the curve   $  T^2-X^2 = \tau^2$  (corresponding to $r = const.$)  until the   intersection of  that curve  with the   curve   $ T_1(X)$.   From there   $\Sigma_\tau $ continues  along the line  $T= const. $    until    the intersection of  this line  with  the curve  $T_2(X)$.  From there it   continues  along the line  $T/X= const. $  (which  corresponds  to  $t = const.$) all the way to the asymptotic  region.

 Finally   we  define  $\zeta$  (on the  region $ X \geq 0 $)   as the {\it distance} of  the  given  space-time point to the  $T$ axis (\emph{i.e.}  the line  $X=0$)  along the hypersurface $\Sigma_\tau $.   For the   region $ X \leq 0 $ we   define  $\zeta$ as    minus  the {\it distance} of  the  given  space-time point to the  $T$ axis  along the hypersurface $\Sigma_\tau $.

 In order to complete the  specification of the new  coordinates and the foliation   all that we need are the  two curves $T_{1,2}(X)$ . They  can be chosen to be
\begin{eqnarray}
T_1(X) = \left(X^2 + \frac{M}{\Lambda^3}e^{-2\Lambda/\sqrt{X}}\right)^{1/2}.
\label{txin}
\end{eqnarray}
The other curve that we need can be found by the  reflection with respect to the horizon $T=X$, of  that given in eq.  (\ref{txin}). This curve is obtained just by interchanging $T$ with $X$ in eq.  (\ref{txin}). That is,   $ T_2$  is   defined   via  the implicit function theorem,   as the solution to the equation:
\begin{eqnarray}
X(T) = \left(T^2 + \frac{M}{\Lambda^3}e^{-2\Lambda/\sqrt{T}}\right)^{1/2}.
\label{xtout}
\end{eqnarray}

It  is clear  from the above recipe that we face smoothness issues   at the   junction points  defining the foliation, and  that this is   problematic regarding the   smoothness of coordinates  we have defined. However  it seems   clear  that  a  simple    smoothing  procedure should  serve to resolve the problem   without affecting the  essence of the construction  (we  will not   further  consider   that aspect   in the  present   work).

It is   clearly important to ensure that these  hypersurfaces   do not cross each other.  We show this in Appendix \ref{proof}.

Essentially, this construction now divides the spacetime into three regions as shown in Fig. \ref{folfigdiv}. Region-A and C are defined, respectively, at the inside and outside of the  event horizon, whereas, Region-B connects them to complete the foliation. Note that due to our choice of intersecting curves eq.  (\ref{txin}) and eq.  (\ref{xtout}), Region-B starts from a point and asymptotically it also ends in a point. In Fig. \ref{folfig} (upper) we plot various space-like hypersurfaces generated by using the above prescription of foliation. As required they ``evolve forward in time"  (in the sense that $\tau$ is  a   good  time  function) and do not cross each other at any stage. An artistic picture of the above scheme in the conformal diagram is also shown in Fig. \ref{folfig} (lower).

\begin{figure}[t]
\centering
\includegraphics[width=10cm, height=5cm]{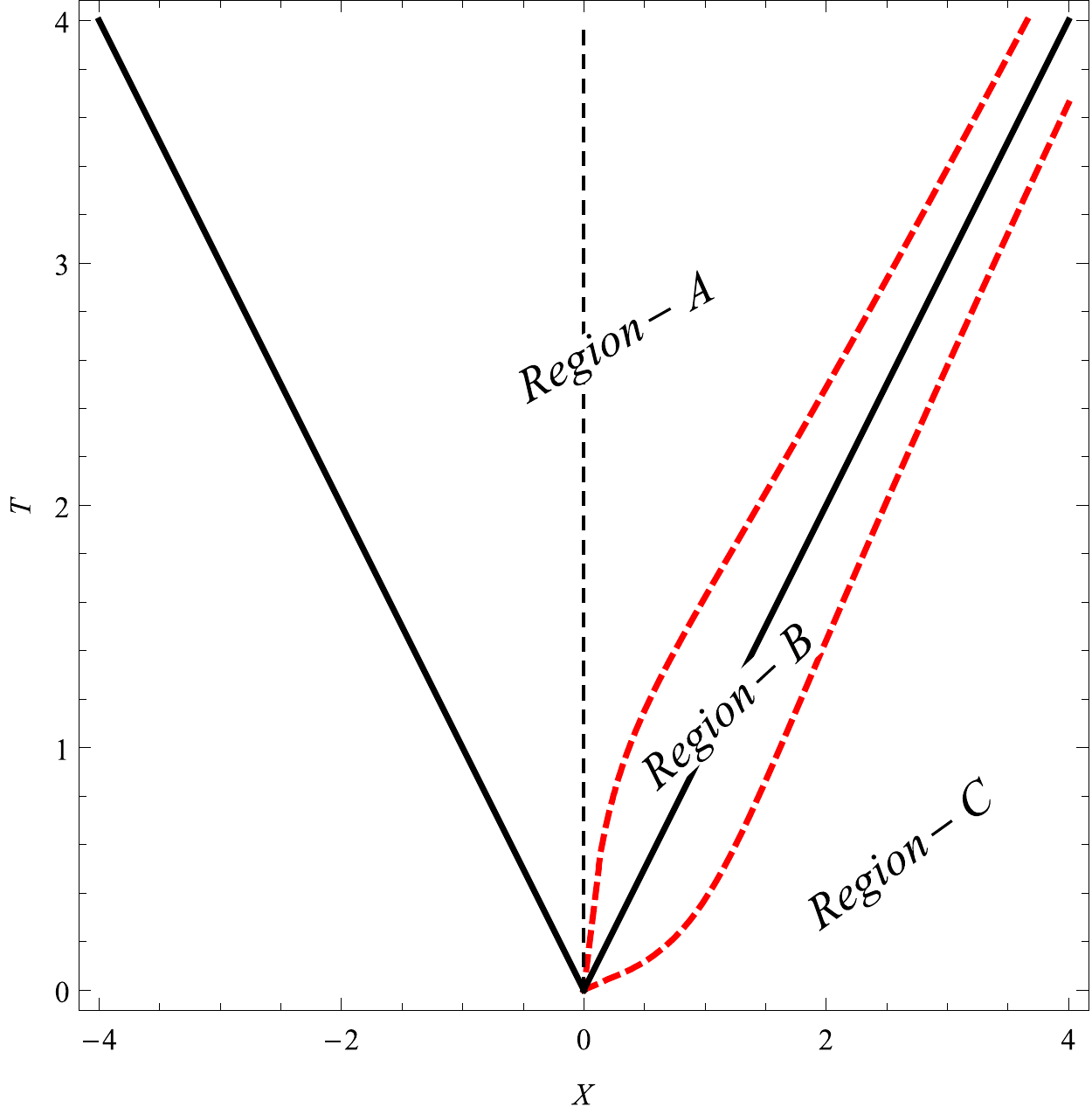}
\caption{Division of the CGHS spacetime in Kruskal coordinates due to the foliation. In this plot we have set $M/\Lambda^3 = 4.42$. For details see the text.}
\label{folfigdiv}
\end{figure}

\begin{figure}
\centering
\includegraphics[scale=0.35]{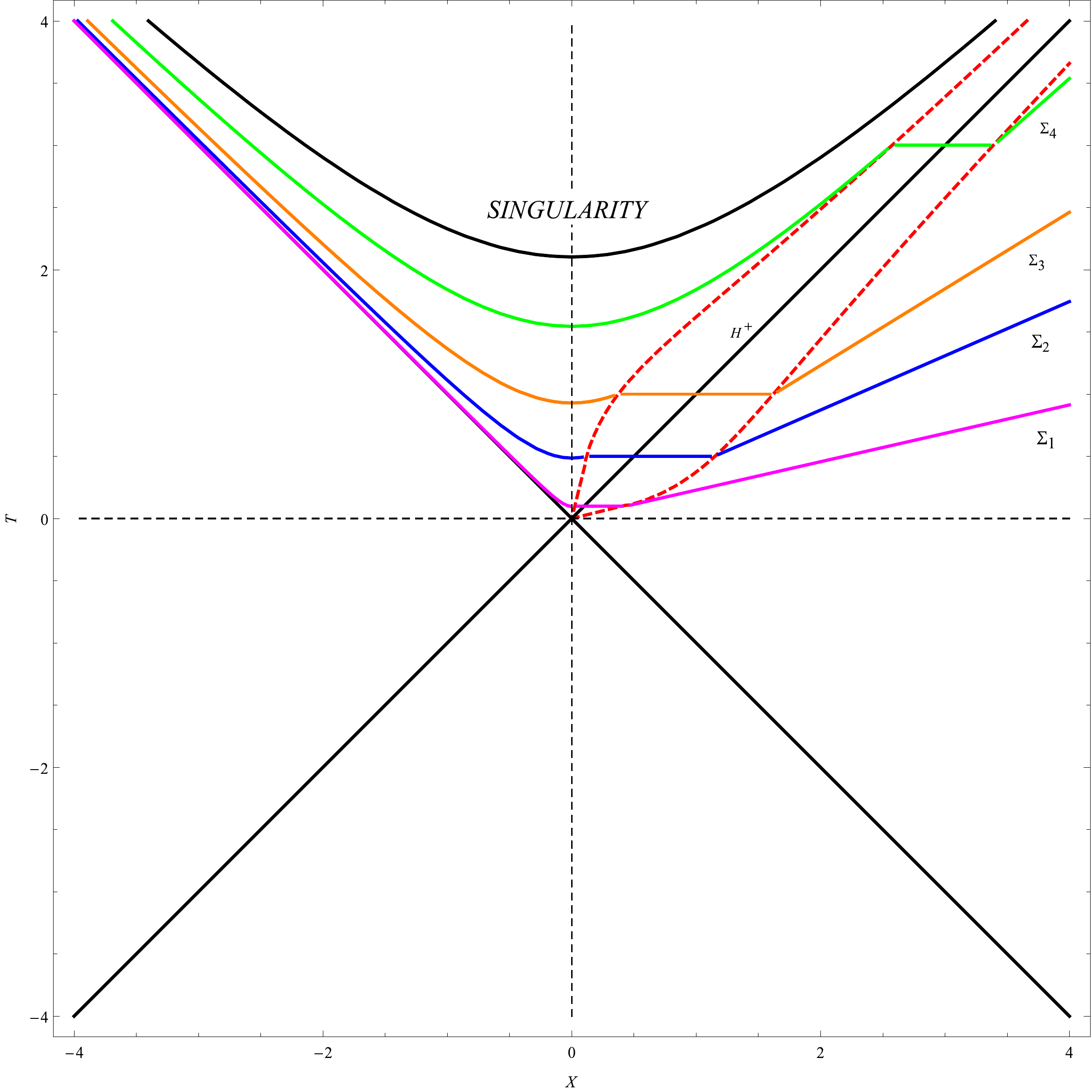}
\includegraphics[scale=0.5]{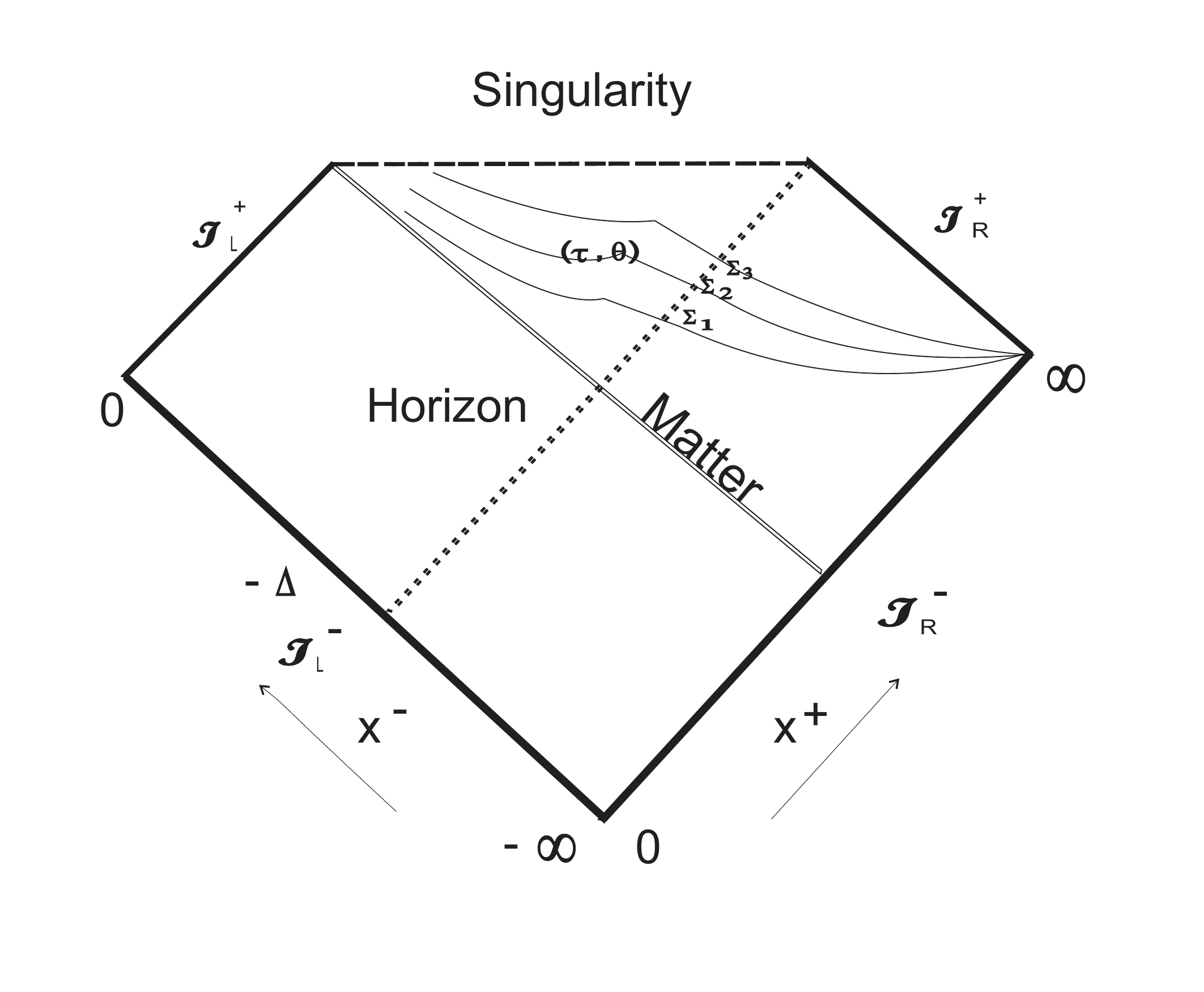}
\caption{Spacetime foliation of the CGHS spacetime. The upper figure (in Kruskal coordinates) contains mathematical plots of various slices with fixed value $M/\Lambda^3 = 4.42$. We have chosen $T= 0.1$ (magenta), $T = 0.5$ (blue), $T=1$ (orange), $T = 3$ (green) for connecting $T=const.$ lines. In the lower figure various foliating surfaces are highlighted in the conformal/Penrose diagram of CGHS model. For more details see the text.}
\label{folfig}
\end{figure}

\begin{figure}
\centering
\includegraphics[width=12cm, height=6cm]{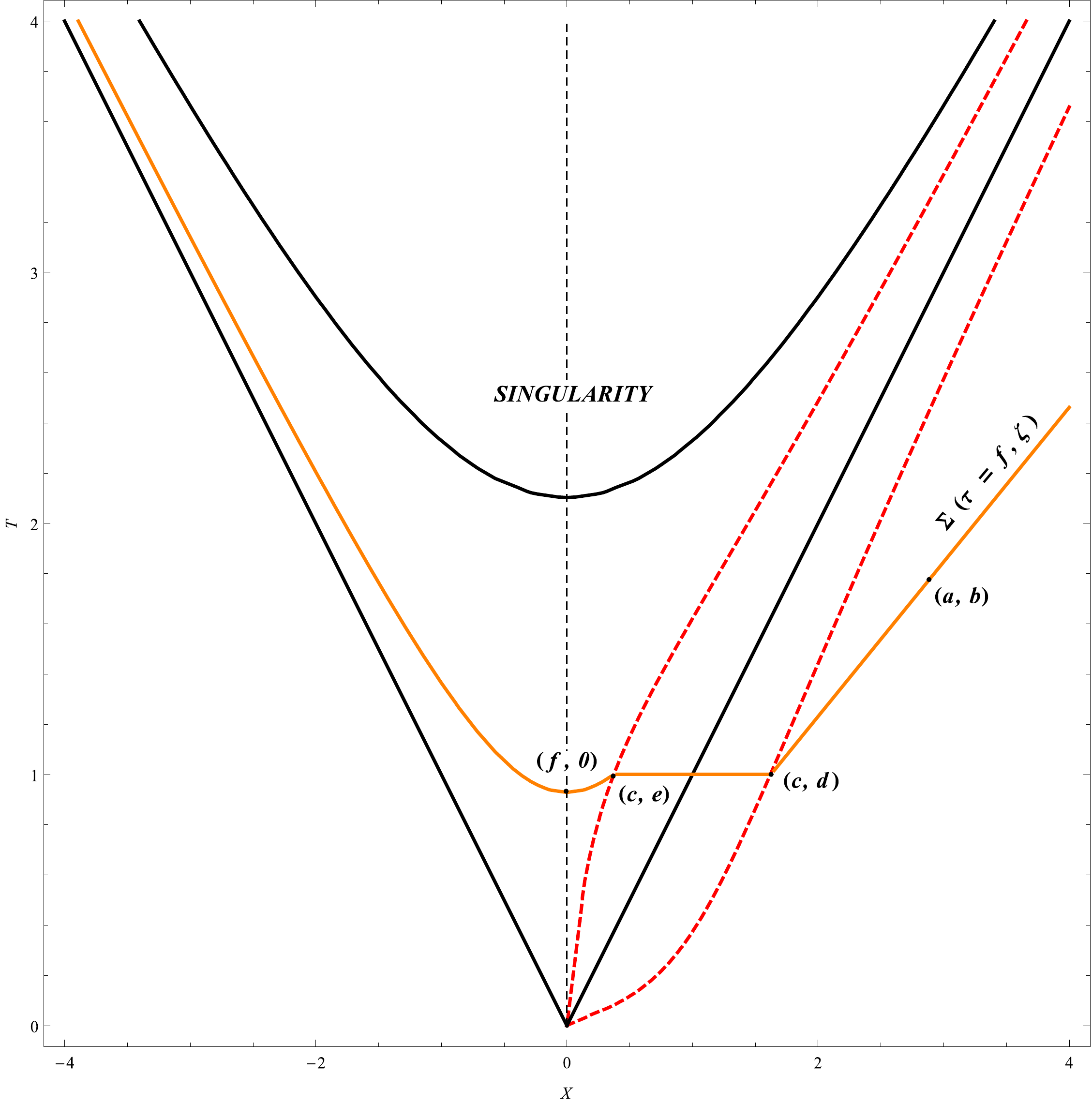}
\caption{ ($\tau, \zeta$) parametrization of spacelike hypersurfaces used to foliate the Kruskal diagram of CGHS spacetime. Like before, these mathematical plots have fixed values of $M/\Lambda^3 = 4.42$. For detailed description see the text.}
\label{param}
\end{figure}

Now,  we give the explicit transformations of coordinates $\tau (T,X), \zeta(T,X)$. For that, let us, for instance, consider the general case of a point in Region-C. All other cases can be easily found using the same recipe. We refer the reader to consult  Fig. \ref{param} to get familiar with the scheme as presented below.

Let's take a point $T=a,~ X=b$ in Region-C. Given these values we have a unique $t$ from eq.  (\ref{tout}). Now,  we follow this line of $t = const.$ until we intersect the curve eq.  (\ref{xtout}). This determines a point $T=c,~X=d$. Then we move along $T=const.=c$ until we intersect the curve eq.  (\ref{txin}). This gives a point $T=c,~X=e$. From this point we move along $r=r(c,e)=const.$ given by eq.  (\ref{rin}) until we intersect the axis $X=0$. This gives us a value of $T=f$ and we take this as the value of $\tau=\tau (a,b)$. In this way we assign a unique value of $\tau$ for the entire hypersurface. The other coordinate $\zeta$ is now the distance between $(T=f,X=0)$ and $(T=a, X=b)$ along the hypersurface which determines $\tau$. This distance is given by
\begin{eqnarray}
\zeta = \zeta_A + \zeta_B + \zeta_C,
\label{distance}
\end{eqnarray}
where
\begin{eqnarray}
\zeta_A = \int_{r=const.}ds_{A} &=& \sqrt{\frac{A}{M/\Lambda - \Lambda^2 A}}\int_{f-\Delta}^{c-e-\Delta}\frac{dx^{-}}{x^{-}+\Delta}, \label{ds1}\\
\zeta_B = \int_{T=const.}ds_{B} &=& \int_{c-e-\Delta}^{c-d-\Delta} \frac{dx^{-}}{\sqrt{{\frac{M}{\Lambda}} + \Lambda^2 (x^- +\Delta) (x^- - 2c)}}, \label{ds2}\\
\zeta_C = \int_{t=const.}ds_{C} &=& \int_{c-d-\Delta}^{a-b-\Delta} \frac{dx^-}{\sqrt{\Lambda^2 x^- (x^- + \Delta) + \frac{M(b-a)}{\Lambda(b+a)} }}. \label{ds3}
\end{eqnarray}
The explicit values of these integrations are provided in Appendix \ref{int}. In a similar manner, for other cases when a point belongs to Region-B or Region-A, one should integrate up to $ds_B$ or $ds_C$, compatible with the limits of integrations.

In the $(\tau,\zeta)$ parametrization the expression of Ricci scalar in Region-A is constant on each $\Sigma_{\tau}$ hypersurface. The explicit expression for $R$ in Region-A follows from eq.  (\ref{ricci}) and eq.  (\ref{rin}) and is given by
\begin{eqnarray}
R (\tau) = \frac{4 M \Lambda}{M/\Lambda - \Lambda^2 \tau^2}.
\label{rtau}
\end{eqnarray}
Note that the position of singularity is now given by a finite $\tau_{s} = \frac{M^{1/2}}{\Lambda^{3/2}}$, and we are interested  in  using this foliation to evolve various quantum states (using CSL) in the open interval $(0,\tau_s)$.

The  foliation can be  continued backwards to cover the rest of the space-time  before the singularity in an arbitrary way, because  its   exact  form  will have    very little  effect on the final  result
of   the  state  evolution  resulting from  the CSL   interaction (recall we   are  using a  kind of  interaction picture and   treating  the  CSL  specific  modifications  of the dynamics   as an  ineteraction)  simply  because in our proposal the CSL parameter    will
only become  large  in the regions of large curvature   which  are precisely  those  covered  by the specific   foliation  presented  above.
 This  assumption  will be   central to  the picture whereby,   information loss in black hole evaporation   is controlled  by the same  process   that, according to collapse theories,    controls  everyday  situations, and  helps  resolve the measurement problem.

\subsection{Specification of collapse operators in modified CSL}
\label{subsec:operators}
 The CSL   theory  can be generalized so  as to drive/collapse  an initial state  into a  joint  eigenstate  of a set of   mutually commuting operators $\lbrace A^\alpha\ \rbrace $. 
This requires the   introduction  of  a white noise function $w^\alpha$ for each  one of  the  $A^\alpha$'s . In this case,  the  equation  corresponding  to eq. \eqref{CSL6} takes  the form:
\beq\label{dm_gen}
\rho(t) = {\cal T}e^{-\int_{0}^{t}dt'\big[i[\underrightarrow{\hat H}-\underleftarrow{\hat H}]+\frac{\lambda}{2}\sum_\alpha [\underrightarrow{\hat A}^\alpha-\underleftarrow{\hat A}^\alpha]^{2}\big]}\rho(0).
\eeq

We call  $\{A^\alpha\}$ the \emph{set of collapse operators}.  In order to adapt CSL evolution to the CGHS scenario, a situation that involves both quantum fields and gravitation, and  as  anticipated  in the previous  subsection, we  will consider that the rate of collapse is enhanced by the curvature of the space-time, so that, as the evolution approaches the singularity (in a finite time) the rate of collapse will diverge.
This  will  ensure that a  complete collapse of the   state   of the field to one of  the  eigenstates of the chosen operators  occurs in a finite  time.
Far from the singularity the rate of collapse will be much smaller,  and thus  the effects of CSL evolution will be negligible. 


Also, for this matter, collapse operators have to be  smooth and locally constructed from the quantum fields.
From these considerations, we choose the collapse operators as the operators that count the number of right-moving \emph{particles inside the black hole} in a definite state, as described by observers in late times (that is, for observers that describe the  Hawking radiation{\footnote{Even  though in regions   (such as  the black hole  (BH)   interior)  where  one  does not have  a time-like Killing field (KF) and thus one   cannot  rigorously  talk   about the   conserved energy of  a   classical particle,  when  one  has  a KF   which  is  time like  in  some   other region, the  conservation law  associated   with it  can  be used to extend the characterization in terms of   ``energy'' of  classical  particles  localized where the   KF is  not time-like. That is  how  one  talks, for instance, about  the    negative  energy of  a  particle inside the  ergosphere of a Kerr BH.   In fact,  conservation laws   even  allow sometimes  the effective  use of the  notion  in order to make  predictions   that  are  relevant  for observers   in the   asymptotic  region.   One  can   thus talk  about the  ``time''  KF inside the  horizon of a Schwarzschild space-time   $ t^a $ (the field  associated  with  Schwarzschild     coordinate $t$  suitably  extended  to the black hole  interior) although,   in this region,  this   KF  is not time-like.  One can use the  behavior of a certain mode of a  quantum field in the black hole  interior  and call it  (in an abuse of language)  its  ``energy''   according to   asymptotic observers that see the  BH  at ``rest''  ( i.e. one can take  the  operator  $  i t^a   \nabla_a  $  and  apply it to the  field  mode  as  a way to characterize it,  even if the mode is fully located in the BH  interior).  This scheme  allows  us to  characterize  the particular modes   we have  used in the BH  interior,  although no further physical significance is being attached to  this  characterization.}}). Recall that the right moving modes of the field inside the black hole are given by eq. \eqref{vint}, and the left moving remain the same as in eq. \eqref{eq.6}.
The Fock space of states of the quantum field in the interior black hole region, $\mathscr{F}^{int}_R\otimes \mathscr{F}^{int}_L$ has as a basis the set $\{\kets{F}^{int}_R\otimes\kets{G}^{int}_L\}$.

The action of the right-moving particles number operator%
\footnote{Using the Klein-Gordon inner product the operator $(N^{int}_{R})_{nj}$ can be expressed in terms of the field $\hat f$ as a product of integrals over some arbitrary Cauchy hypersurfaces  as:
\begin{multline}
(N^{int}_{R})_{nj} =  \hat{\tilde{b}}^{R\,\dagger}_{nj} \hat{\tilde{b}}^{R}_{nj}= \int_\Sigma d\Sigma \int_{\Sigma'} d\Sigma' n^\mu n^{\nu\prime} \times \\
\times \Big(
\tilde{v}_{nj} \tilde{v}_{nj}^{*\prime} \nabla_\mu \hat f^\dagger  \nabla_\nu \hat f'
-\tilde{v}_{nj}( \nabla_\nu \tilde{v}_{nj}^{*\prime}) (\nabla_\mu \hat{f}^\dagger ) \hat f'
-\tilde v_{nj}^{*\prime} (\nabla_\mu \tilde{v}_{nj} ) \hat f^\dagger \nabla_\nu \hat{f}'
+ (\nabla_\mu \tilde v_{nj} ) (\nabla_\nu \tilde v_{nj}^{*\prime} ) \hat f^\dagger \hat f'
\Big)
\end{multline}
where, in an abuse of notation, we have dropped the superscript $R$ on the modes. The \emph{out} field operator $\hat f$ is defined in eq. \eqref{fout} and the modes have been discretized as in eq. \eqref{ulr}. Note, however, that the $\tilde{v}_{nj}$ modes have support only inside the horizon.%
}
$(N^{int}_{R})_{nj} =  \hat{\tilde{b}}^{R\,\dagger}_{nj} \hat{\tilde{b}}^{R}_{nj}$ acting on  $\mathscr{F}^{int}_R$ is the following:
\beq
(N^{int}_R )_{nj} \kets{F}^{int}_R= F_{nj} \kets{F}^{int}_R.
\eeq
The set of collapse operators we are proposing  for the modification of CSL is  $\{\tilde{N}_{nj}\}$, where
\begin{equation}\label{Ntilde}
\tilde{N}_{nj} \equiv (N^{int}_R)_{nj} \otimes \mathbb{I}^{int}_L\otimes \mathbb{I}^{ext}_R\otimes \mathbb{I}^{ext}_L
\end{equation}
where $\mathbb{I}^{int}_L$, $\mathbb{I}^{ext}_R$, and $\mathbb{I}^{ext}_L$ are  the identity operators in the corresponding Fock spaces. Any state in the basis of $\mathscr{F}^{out}$ (see eq. \eqref{Fout}) is now an eigenstate of the collapse operators $\tilde{N}_{nj}$:
\begin{equation}
\tilde{N}_{nj} \kets{F}^{int}_R \otimes\kets{G}^{int}_L \otimes \kets{H}^{ext}_R \otimes \kets{I}^{ext}_L=F_{nj} \kets{F}^{int}_R \otimes\kets{G}^{int}_L \otimes \kets{H}^{ext}_R \otimes \kets{I}^{ext}_L.
\end{equation}

As  we   already  noted, the states characterized by a definite number of particles in  the individual modes we  are working with, are not of the  Hadamard  form  and they  correspond to  a  singular   behavior  of the energy-momentum tensor   expectation value,   precisely along the  black hole  horizon.    However,  we  must  note that the standard  CSL  evolution  controlled  by  a fixed  value of $\lambda$ is   meant to  be a smooth one  and,  in fact,    it only drives  the state to one of the  eigenstates  of the  collapse operator  in the limit  $ t \to \infty$.

In our case, as  we  already anticipated,  we   will choose   a variable  but  smooth   $\lambda$  which, however,  will be bounded   except  as one  approaches the singularity where,  one expects,   quantum gravity  effects  to    become dominant,  and  where   the   space-time picture  will likely break down.    Moreover,  in  our case   we   will  chose  the   CSL  evolution  that is  driven   by collapse operators that   are  smooth  and  locally  constructed   from     the  quantum fields,   the  initial state  of the  field  (as defined  in   past null infinity  as in eq.  \eqref{psi_ini}) is  a regular  (Hadamard)   state  with  a   smooth  expectation value for   the  energy-momentum tensor,  and  thus  the   state resulting  from  the  evolution at  any   hypersurface   of our  foliation   lying  before the  singularity,   will   also  be a   regular  (Hadamard)  state  with a   smooth expectation for the energy momentum  tensor. Again this is   just as  it  occurs  in the case of  the state resulting  from  the Minkowski  vacuum of  a quantum  field  interacting with a  physical detector which  is  characterized  by  a  smooth  operator and  thus, as  we discussed, can not  measure  the precise  number of Rindler particles in a  particular mode in a finite  amount of time.

\subsection{The curvature dependent coupling $\lambda$ in modified CSL}
\label{subsec:lambda}
As we have said, in the toy model of CSL we are proposing the rate of collapse is enhanced by the curvature of the spacetime. This mechanism is introduced in terms of the rate of collapse $\lambda$, which will be dependent, in this case, on the Ricci scalar:
\begin{equation}\label{lambdaR}
\lambda (R) = \lambda_0 \left[1+\left(\frac{R}{\mu}\right)^\gamma\right],
\end{equation}
where $R$ is the Ricci scalar of the CGHS space-time and $\mu$ and $\gamma \ge 1$ are constants.

This is a key hypothesis made in the paper, which states that there is some   connection between the physics that    underlies the     dynamical collapse of quantum states with gravity. The argument is that assuming the rate of collapse intensifies in a region of high curvature, information might be completely destroyed due to the stochastic nature of this process. Of course, this hypothesis cannot be confirmed or ruled out as of now.  In order to do that, one would   need to perform experiments or analyze observations of a quantum process in presence of ultra-strong gravity{\footnote{If it turns   out to be the case that strong curvature expedites quantum collapse of wave function, it will of course be a strong evidence    in favor of our proposal that information is indeed lost in black hole evaporation in the manner we have   exemplified with this example. This in fact makes our proposal, in principle,    susceptible to empirical investigation and that is an advantage over  various models that have been   put forward in this context so far.}}.

Note that the hypersurfaces given by the foliation  in subsection \ref{subsec:foliation} have constant $R$ inside the black hole (in almost all the part of $\Sigma_\tau$ that lies inside). Then, from eq. \eqref{rtau} we have that for the region of interest, the rate of collapse depends on the time parameter defined by the foliation, $\lambda=\lambda(\tau)$. Standard CSL is defined for a constant value of $\lambda$, however, the generalization to time-dependent $\lambda$ is simply done by making the substitution $\lambda\to\lambda (\tau)$.

This  choice for  the curvature dependence of the CSL  parameter  is meant to ensure that  the initial  state would be  driven to  an  eigenstate  of the  collapse operators in a finite  amount of  {\it time}  $\tau $   so  that,  if we  were to  continue the  evolution  up to  the singularity (we  will not contemplate  that  simply  because,  as  we indicated,  it is  within this region where we expect quantum gravity  effects  to    dominate  and   the   space time picture   to  break down,  invalidating most  of our considerations), the state of the  field  would be one with definite number of particles in each mode.  The point, however, is that such  (singular)  state    would  only  be approached   as  one approaches the singularity, and the state of the field,  in the region before the  singularity  where  semiclassical  considerations  are   expected to hold,  will  be  a state  with  a   smooth  expectation value for the energy-momentum tensor.

At this point, we must  comment on the importance of the dimensionality of the model under consideration in selecting the curvature dependent coupling of the parameter $\lambda$. Clearly, for a more realistic, four dimensional models, this choice has to be done in a way that it does not lead to big deviations in the evolution of the early Universe. Otherwise such changes would have to be carefully investigated. In making our proposal we take Penrose's Weyl curvature hypothesis \cite{weyl-penrose} as a guiding principle. As already mentioned in the earlier work \cite{Okon2}, in a four dimensional model, $\lambda$ can be naturally taken to be a function of Weyl scalar $W_{abcd}W^{abcd}$ and, as the early Universe is thought to have a comparatively small value of such quantity, the deviations from the standard quantum mechanical consideration can be expected to be insignificant. However, for the two dimensional case at hand, Ricci scalar can be chosen to be the only one algebraically independent component of the Riemann-tensor, and thus we take $\lambda$ to depend on $R$ as defined in \eqref{lambdaR}.

\subsection{Initial state and modified CSL evolution}
\label{subsec:csl-evol}

As we have mentioned in section \ref{sec:CGHS}, a left-moving pulse of matter produces the CGHS black hole. Then, the initial state of the quantum field, defined in $\mathscr{I}^-$, that will evolve in this space-time is
\begin{eqnarray}\label{psi_ini}
\kets{\Psi_{i}} & = &\kets{0}^{in}_R\otimes\kets{Pulse}_L \\\nonumber
   &=& N \sum_F e^{-\frac{\pi}{\Lambda} E_F} \kets{F}^{int}_{R}\otimes\kets{F}^{ext}_{R} \otimes \kets{Pulse}_L \,,
\end{eqnarray}
where we have used eq. \eqref{invac}. The state $\kets{Pulse}_L$ can be considered as a very localized left-moving wave packet. Because of the dependence of the CSL parameter $\lambda$ on the curvature we can assume that the state $\kets{\Psi_i}$ will remain mostly unchanged as it evolves outside the horizon since $\lambda$ would be very small (recall we are working in the interaction picture) until it reaches some hypersurface $\Sigma_{\tau_0}$ described by the foliation given in section \ref{subsec:foliation}.
In this section  we derive the final density matrix for the modified CSL evolution of the initial state $\kets{\Psi_{i}}$, eq. \eqref{psi_ini}, from the initial hypersurface $\Sigma_{\tau_0}$ to a final hypersurface $\Sigma_{\tau}$.

The final state of this evolution will depend on the particular set of stochastic functions $w^\alpha$ that occurred during the evolution. In order to take into account our ignorance on this particular realization of the $w^\alpha$, we will describe the evolution by considering an appropriate ensemble. So let us consider an \emph{ensemble of systems} identically prepared in the same initial state $\kets{\Psi_{i}}$\footnote{ At this  point one  could  imagine   instead  just one  system  in a  given initial state, and consider  the ensemble of  all alternative evolutions  as represented  by  the density matrix.    Although there  is,  in  principle,  nothing wrong with  such view,  we  will   not adopt  it here,   as   we  feel   it  could  more easily  lead  to  conceptual confusions.}. This ensemble is described by the pure density matrix $\rho(\tau_0) = \kets{\Psi_i}\bras{\Psi_i}$. Note that even though we are talking about an ensemble, the density matrix is pure because all systems are in the exact same state. The evolution of this density matrix can be done simply using CSL.

Then, for the free field evolution,  eq. \eqref{dm_gen} reads:
\beq\label{dm1}
\rho(\tau) = {\cal T}e^{-\int_{\tau_0}^{\tau}d\tau'\frac{\lambda(\tau')}{2}\sum_{nj} [\underrightarrow{\tilde N}_{nj}-\underleftarrow{\tilde N}_{nj}]^{2}}\rho(\tau_0).
\eeq
Note that  $\rho(\tau_0) = \kets{\Psi_i}\bras{\Psi_i}$ can be expressed as:
\begin{equation}
\rho(\tau_0) = \kets{0}^{in}_R\bras{0}^{in}_R \otimes \kets{Pulse}_L \bras{Pulse}_L = \rho_R (\tau_0) \otimes \kets{Pulse}_L \bras{Pulse}_L .
\end{equation}
The evolution operator in eq. \eqref{dm1} acts only on $\rho_R (\tau_0)$. The right-moving initial density matrix, $\rho_R (\tau_0)$ takes the following form when expressed in terms of the \emph{out} quantization:
\beq
\rho_R (\tau_0) = N^2 \sum_{F,G} e^{-\frac{\pi}{\Lambda} (E_F +E_G )}
\kets{F}^{int}_R\otimes\kets{F}^{ext}_R \bras{G}^{int}_R \otimes\bras{G}^{ext}_R  .
\eeq
On the other hand, we have
\begin{multline}
\sum_{nj} [\underrightarrow{\tilde{N}}_{nj} - \underleftarrow{\tilde{N}}_{nj}]^2
\kets{F}^{int}_R\otimes\kets{F}^{ext}_R \bras{G}^{int}_R \otimes\bras{G}^{ext}_R 
 = \sum_{nj} (F_{nj}-G_{nj})^2
\kets{F}^{int}_R\otimes\kets{F}^{ext}_R \bras{G}^{int}_R \otimes\bras{G}^{ext}_R .
\end{multline}
 The calculation is  facilitated by the fact  that the $\tilde N_{nj}$ and their eigenvalues  do  not depend on  $\tau$. Thus, for any $\tau$, we have:
\begin{equation}\label{rhot}
\rho_R (\tau) = N^2 \sum_{F,G} e^{-\frac{\pi}{\Lambda} (E_F +E_G )} e^{- \sum_{nj} (F_{nj}-G_{nj})^2\int_{\tau_0}^{\tau}d\tau'\frac{\lambda(\tau')}{2}}
\kets{F}^{int}_R\otimes\kets{F}^{ext}_R \bras{G}^{int}_R \otimes\bras{G}^{ext}_R.
\end{equation}
In general, this  density  matrix does not represent a thermal state. Nevertheless, as $\tau$ approaches the singularity, say at $\tau=\tau_s$, the integral{\footnote{The integral is carried out assuming $\gamma =1 $ in eq.  (\ref{lambdaR}). However,  one can do it for  other  values of $\gamma \ge 1$ with a similar conclusion.}}
\begin{eqnarray}
\int_{\tau_0}^{\tau}d\tau'\lambda(\tau') = \lambda _0 \left(\frac{4 \sqrt{M\Lambda } \tanh ^{-1}\left(\frac{\Lambda ^{3/2}}{\sqrt{M}}\tau\right)}{\mu }+\tau \right)_{\tau_{0}}^{\tau_s = \frac{\sqrt{M}}{{\Lambda ^{3/2}}}},
\end{eqnarray}
 diverges simply because   of  the  way   $\lambda(\tau)$  increases with  curvature. Therefore, as   $\tau\to\tau_s$ the non diagonal elements of $\rho(\tau)$  vanish and we have in this limit:
\beql{rhot2}
\lim_{\tau\to\tau_s} \rho_R (\tau) = N^2 \sum_{F} e^{-\frac{2\pi}{\mu} E_F}  \kets{F}^{int}_R\otimes\kets{F}^{ext}_R \bras{F}^{int}_R \otimes\bras{F}^{ext}_R.
\eeq
Thus, when $\tau\to \tau_s$, the complete density matrix that represents the evolution of state eq. \eqref{psi_ini} is given by
\begin{equation}
\lim_{\tau\to\tau_s} \rho (\tau) = N^2 \sum_{F} e^{-\frac{2\pi}{\mu} E_F}  \kets{F}^{int}_R\otimes\kets{F}^{ext}_R \bras{F}^{int}_R \otimes\bras{F}^{ext}_R \otimes \kets{Pulse}_L\bras{Pulse}_L.
\label{rhot3}
\end{equation}
Note that $E_F$ represents the energy of state $\kets{F}^{ext}_R$ as measured by late time observers. The operator given by eq. \eqref{rhot3} represents the ensemble when the evolution has almost reached the singularity (at the hypersurface $\Sigma_{\tau\to\tau_s}$). However, we want to describe the evolution beyond the singularity for which we have to consider the effects of a plausible quantum theory of gravity. We elaborate on this matter in the next subsection.


\subsection{A task for quantum gravity}

    The fact that  CSL evolves   states towards    one of the  eigenstates of the collapse operators ensures that as  the result of   our evolution,  the   state  at   hypersurfaces  $\tau$ constant but  very close to the  singularity     would   be  of the form
  \begin{equation}
\kets{\Psi_{\tau\to\tau_s}} =  \kets{F}^{ext}_R \otimes \kets{F}^{int}_R\otimes \kets{Pulse}_{L}
\end{equation}
for some particular number of particles distribution $F$.
Note that   in the   expression above  there is no  summation  implied over the $F$. The  state is  a pure one,  with  a definite  occupation number  in each  of the field's modes.

 Next, we   consider the  role that  is reserved  for a  quantum  theory  of gravity: We will assume that  a  reasonable  theory  of QG     will resolve the singularity  and   lead,  on  the other  side, to   a reasonable space-time. Moreover, we  will assume that such a theory  will not lead to  large violations  of the   basic  space-time conservation laws.

On the basis  of  these  assumptions,  we  consider  the situation  on the  hypersurface  that lies  ``just   before the   singularity" (we could  define such hypersurface more precisely  by requiring that the  curvature reaches a particular value close to that where quantum gravity effects  should become important, say, half the Planck scale).  There,  from the  ``energetic  point of view",  we  would    have  the  following:

{ i)}  The  positive energy flux associated   with  the incoming    left moving pulse  that  formed the  BH.

{ii)} The   negative  flux   associated    with   the  remaining   left  moving  modes  in  a vacuum  state  (from Region-II to Region-III in Fig. \ref{cghs})  which is known to  correspond  essentially   to the  negative  of the  total  Hawking radiation flux (see sub-section \ref{sec:enfl}).

{ iii)}  The flux associated  with the  right moving modes   which crossed  the   collapsing matter and  which    went   directly into the  singularity (from Region-I' to Region-III in Fig. \ref{cghs}).

The only part  missing   in the   above  energy budget  is that  corresponding  to  the Hawking radiation flux.  If, as  we  have assumed,  energy  is to be  essentially  conserved  by  QG,  it seems that the only  possibility   for the  state   in the  post singularity region is  one corresponding to  a very small value of the energy.  It is   possible  that such state  of  negligible energy would  be  associated with a small  amount of   remnant   radiation, or  perhaps some  kind of   stable   Planck  mass remnant.Those  possibilities   would represent a  situation where some  of  the initial  information   survives  the  whole process. For simplicity, we  will ignore that alternative, and  replace it by the simplest thing: a zero  energy momentum  state  corresponding to a region of space-time that  would   necessarily be  trivial.
 We  denote it by {$\kets{0^{post-singularity}}$}.

  Therefore,   we can  complete  the   description  of the  evolution of the initial situation,   by   assuming that  the effects of   QG can be represented  by:  a)  the cure of the singularity and b)  the   transformation of  the state    just   before to that  one   just  after the   would  be  singularity (or  more precisely, the quantum gravity region) represented  as s  follows;

 \begin{eqnarray}
\kets{\Psi_{\tau\to\tau_s}} &=& \kets{F}_R^{ext} \otimes \kets{F}_R^{int}\otimes \kets{Pulse}_{L}\\ \nonumber
&& \longmapsto  \kets{F}_R^{ext} \otimes \kets{0^{post-singularity}}. \nonumber
\end{eqnarray}

 The above result  is a bit unsettling because  we  end up  with a  pure quantum state, but     we do not know  which one.
  That selection,  according to CSL,    is  determined  by  the particular realization of the  collection of random functions $w^\alpha$    that   appear in the    evolution  equation  for the state. That is,  it depends  on precisely the  stochastic  aspects of the  CSL   evolution of the quantum state.

One might be   concerned  about  the naturalness   of the  choice for  the   collapse  operators   used to drive the  CSL   modifications. It   might  be argued  that   it   does   not seem  to be  a   very natural   one  as the notion of particle  number in the interior  region  is  completely  {\it ad hoc}  because   there  is   not even   a time-like  Killing   field.   One can  argue that the  choice  is  made   because   of  the correlations  that occur   in the   in-vacuum,   between  those  states   and  the states  with  well defined energy  in the   asymptotic  region,   which are  states  with a definite  number of  particles in   the outside    region.   We  will not  attempt to  further  justify  such  choice  here,  and  will  only note the relative robustness  of the  analysis that is expected to  emerge  from the fact that  the  density matrix    characterizing a  subsystem of a system   in a pure state (resulting  from  tracing  the rest of the system's  degrees of  freedom) is  independent of the  choice of basis. In fact, as  we  will  see in the next section,  the final result we   obtain  characterizes  the end point  of the black hole  evaporation, in terms  of  a  density matrix. We  will  mention some  subtleties  in this regard after that discussion.

\section{The final result} \label{sec:final-result}

Now  we are ready   to   proceed to the  effective  description of the end  state  taking  into  account the effects of a quantum theory of gravity. In section \ref{subsec:csl-evol} we have already described the evolution up to a hypersurface ``just before the singularity''. The ensemble at this hypersurface  ( $\Sigma_{\tau_s-  \epsilon}$)   is described by the density operator
\begin{equation}
\lim_{\tau\to\tau_s} \rho (\tau) = N^2 \sum_{F} e^{-\frac{2\pi}{\mu} E_F}  \kets{F}^{int}_R\otimes\kets{F}^{ext}_R \bras{F}^{int}_R \otimes\bras{F}^{ext}_R \otimes \kets{Pulse}_L\bras{Pulse}_L.
\label{rhot32}
\end{equation}

\begin{figure}
\centering
\includegraphics[scale=0.35]{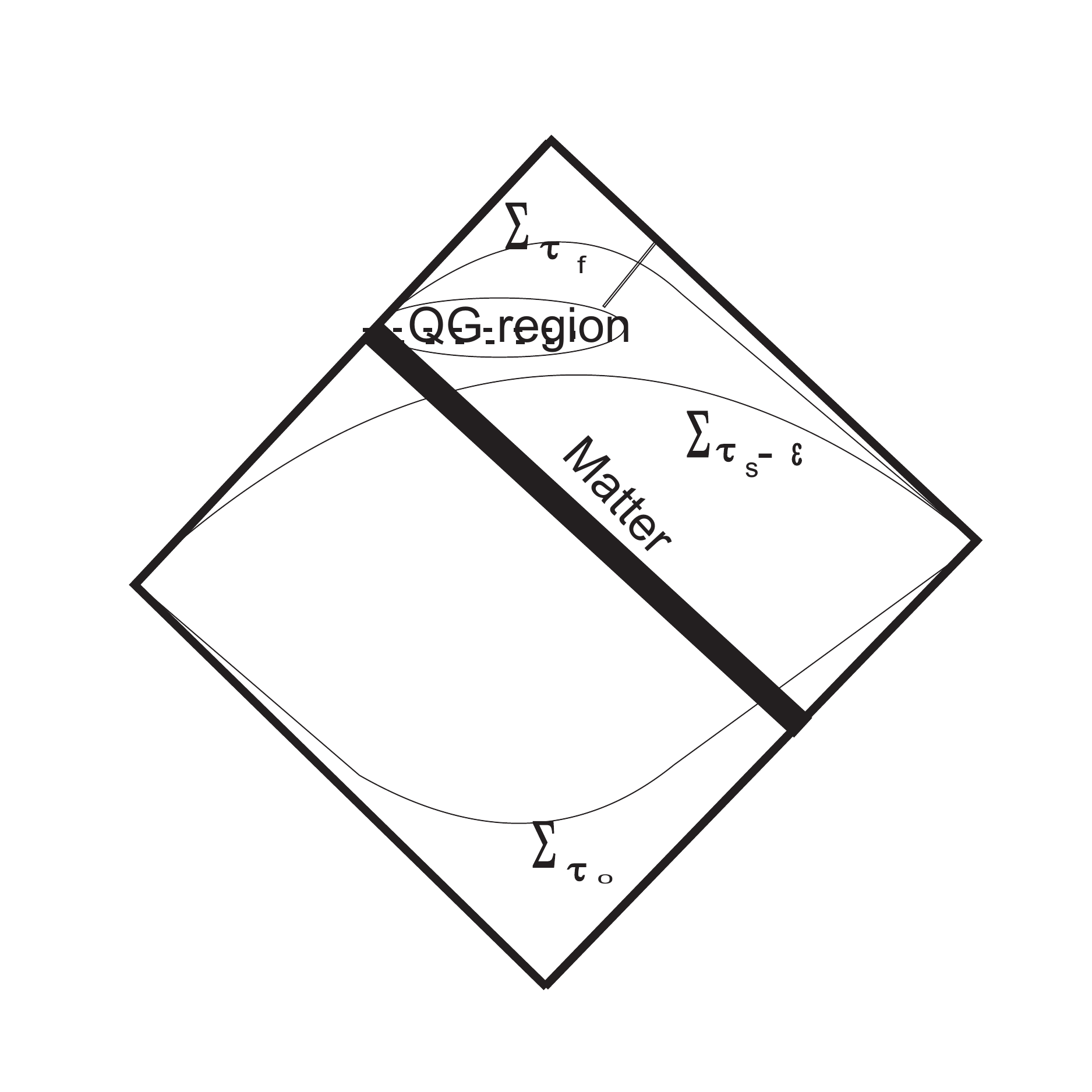}
\caption{Plausible spacetime structure   including  the post-singularity  region. For details see the text.}
\label{finalst}
\end{figure}

If we now  take into account what we  have assumed about QG, we end up with a density matrix  characterizing the ensemble after the  region  that would have corresponded to the singularity,   (i.e   for instance on the  hypersurface   appearing in Fig. \ref{finalst} with no name),  which is given by
 \begin{eqnarray}
 \rho_{final} & = & N^2 \sum_{F} e^{-\frac{2\pi}{\Lambda} E_F} \kets{F}^{ext}_R \otimes \kets{0^{post-sing}} \bras{F}^{ext}_R \otimes \bras{0^{post-sing}} \\ \nonumber
& = &  \kets{0^{post-sing}} \bras{0^{post-sing}} \otimes \rho^{ext}_{thermal}.
\end{eqnarray}

That  is,   we  started  with an initial   pure state  (characterizing   the ensemble of identical  systems  in the  same state)   of the quantum field   corresponding to  an initially  collapsing pulse,   which  we can imagine  as  being   given together   with  the initial  space-time   data on  past  null infinity,  and   we   ended  up with a situation where the ensemble is  characterized  by a \emph{proper mixture}   with  thermal characteristics    on  the early part  of   future null infinity  and    followed  by an empty region. That is   precisely  what was     needed    in order  to reconcile    the Hawking evaporation of a black hole   with Quantum Theory.  Of course in  this  case,   the  result  arises  from    the modified  quantum  evolution provided  by CSL.

   Here,  we   see that   the    questions  concerning  the  choice for  the   collapse  operators   used to drive the  CSL   modifications,  which are  connected  to the  details  of the   collapse
   taking place  in the  inside  region   should   have essentially no  effects  on the overall result  simply  because, as a result of  the  quantum  gravity  resolution of the singularity,   and  the
   considerations  regarding overall energy conservation,    the  details  of  the    state characterizing the field   just  before the region   requiring an essential quantum  gravity   description (\emph{i.e.}, the  ``would be singularity")   are  simply  erased. That is,   as  argued
above,   the   crossing of the ``would be singularity"  should   lead   to:
\begin{equation}\label{post-sing}
\kets{F}^{ext}_R \otimes {\kets{Pulse}}_{L}
 \longmapsto  \kets{0^{post-singularity}}.
\end{equation}
In fact,  if we  concern ourselves    with  the quantities  that  are   conceivable measurable  by    observers  in the outside,  the  issue of exactly  what  operators  drove the collapse in  region III, would be  analogous  to   Bob's  question regarding  the   orientation   of the spin    that was  measured  by Alice on an EPR  situation\footnote{We are  of course  referring to  an EPR  situation  where a   $j=0$   system decays into two     spin $ 1/2$ particles along a  certain axis  and  one   considers    two  observers, Alice  and Bob  who can decide  to measure the   projection  of   the spin of the particle that reaches   each  one, along  various  different  directions.}. That is, the answer can not be  determined  by any measurement     conceivably made  by  Bob  without  input of  regarding what  Alice  does.

\section{Discussions}\label{sec:discussions}

We  have  presented  a  proposal to address the   issue  known as the ``Black Hole Information Paradox''.   The picture   that emerges  is represented in  Fig. \ref{finalst},  which is
essentially the same as    the one presented in \cite{Ashtekar:2008jd};   however, in our  proposal  the  answer   to the  question about  the   fate   of   information is  completely opposite  to that   advocated in that work.    According to the picture  we  have presented  here,   the state   of the quantum   field (corresponding to  the  ensemble of   identically  prepared systems)  on a  very
early  hypersurface,   such as $ \Sigma_0$   (in Fig. \ref{finalst}),  is  pure,   and  then  evolves    according to   CSL resulting   at   late  hypersurfaces    (which  are still to the past of the singularity),   such as  $\Sigma_{\tau_s -\epsilon}$,  in  a   mixed   state (representing,  as   discussed  in section 4,  an  ensemble  of systems)  that  is  in fact  thermal,  according to the set of  asymptotic  observers of the ensemble of evaporating black holes.  That  state corresponds to  a situation where essentially  all the  initial energy (corresponding to the  matter  pulse  in $ \Sigma_0$)    appears  as    radiation crossing the   $\Sigma_{\tau_s -\epsilon}$ hypersurface towards the    asymptotic  region.
 Thus,  when considering the   state  on  $\Sigma_{\tau_s -\epsilon}$, the  information has  already been lost, and the state is  non unitarily  related to that in  $ \Sigma_0$.
  A theory  of   quantum gravity      must  then be evoked  as  resolving the singularity,  and leading   to  a   state of   the matter fields   containing essentially no energy. That is,   the   state of the
  fields   in the region  just   after the    singularity must be  a sort of local vacuum corresponding to a  flat   region of  space-time (\emph{i.e.},  the effects of  quantum gravity must be  represented
  by eq. (\ref{post-sing})).  Then,  on  $   \mathscr{I}^{+}_{R}$,  we  will have   a situation
  corresponding to  a thermal state  in the early  times,    a   small pulse    containing  very little energy  and   associated  with the  start  of  the shadow
    of the  would be  singularity (\emph{i.e},  the boundary  of the causal future of the region  containing the singularity),  a  point  where there  could be   some  influence of  a complex  back reaction,  and perhaps also  some  traces of   influences of quantum gravity, and to the future   of that  point,    the state   is just   the   vacuum.  To  make the  point   more forcefully,  Fig.  \ref{finalst} also depicts  a hypersurface  (with no  particular name)  that encodes  everything  that is  left  after the singularity disappears.  The  hypersurface  has  no points  to the past of the   quantum gravity region that  replaces the singularity.  On  such hypersurface the  state of the field  represents  the Hawking  radiation and  nothing else.    According to this picture,  during the  black hole  evaporation information is   lost, however,  that   is   just the   result of  the modified  evolution described   by   a dynamical reduction    theory  such as CSL (which is  meant to  be the   correct theory that   replaces ordinary  quantum mechanics).  In other  words,   information is    lost,    in this case,    just as  it  is  lost   in all physical process,  with the only difference  being the   effective rate for this   loss  which,  in the case of the  black hole,  is  enhanced  by  our  hypothesis regarding   the dependence of   $\lambda $ on  curvature.

There are,  of course,  many issues that are  not   completely  satisfactory in the     specific account  given   throughout this   work and shortly we shall elaborate on them. However, we feel that the  essential picture    would remain valid in  more realistic  treatments.  To start  with, the    example which  we worked  is   two dimensional while   our  space-time  has   4 dimensions.     We  do not think  that  this  poses a   serious  problem. Among others, one issue that  concerns  people  regarding  the notion of   information loss,  are the  arguments that any such process  must  be  accompanied by the unacceptable breakdown of either, causality, or energy-momentum conservation \cite{bps}. Such arguments  have  been confronted   in \cite{Unruh-Wald},  by  providing    examples   of  evolution laws capable of leading  from pure states to mixed states  under conditions  involving Planck scale physics,  while  locality and causality  are preserved,   and  where  the deviations from ordinary dynamics are exceedingly  small, and  thus  empirically acceptable  for  laboratory situations.

As  we explained, the  BHIP   appears   if  one  assumes that    quantum gravity   resolves  the BH  singularity, and, therefore, removes the need to incorporate  an additional space-time  boundary.  One  might  argue that, even if quantum gravity  resolves the singularity,  one  should  still  keep  an additional boundary  where information  is  registered. One  might, for  instance,   take the  view  that quantum gravity itself leads to the  loss  of information and that  an  extra  boundary has  to  be  added  in  any   description of  a  classical space-time,  if    it is  known  that   there is a non-classical  region, that can only be characterized in  terms of  a quantum gravity language.   The boundary could then be thought as the receptacle for the  apparently  lost information, and  the  inclusion and characterization  of  the quantum  state there, as  restoring  a fully unitary evolution.
One   fact  that  suggests that  it is  worthwhile   looking for  a   different solution   from  the  one considered  above,   is that the hypersurface where  the lost    information is   supposed to be encoded  ({\emph{i.e.}},  that representing the new  boundary of space-time),  would have to be   contained in  a region where, as  a result to the  negative Hawking flux, there  should be   very  little or no energy remaining   at all.  Thus,  if we  were to  adopt  such a view,   we   would  be  considering the  near singularity  region, and whatever  replaces the singularity  in  a theory of quantum gravity,     as the  region    that   contains   the missing   information, despite the fact   that   such  region  must be deemed to contain  essentially no  energy.   That is,  even   before considering the singularity,  we   can think   that there  is  a problem   similar  to  that  in the  standard  BHIP,   simply because  an enormous  amount of information must  be contained in a region having essentially no  energy.  The scheme we  have   been considering  in this manuscript  clearly eliminates also that   aspect  of   the problem.

    Furthermore,  in taking a view like  the last one  considered above,     one  ends  up acknowledging that    the quantum gravity effects that    resolve the singularity are,  at the same time, responsible  for the information loss.
    The idea   would be then,    that  such  information loss  occurs only in connection  with   regions that  would  correspond to classical  singularities,   and   which  we  normally associate  with  very    special conditions     such as  those  connected   with black holes. However,  at the fundamental level,  one  would have to say that  a theory of quantum gravity,    which is   in principle
    able to   lead to information loss, also   underlies  our  notions  of space-time in  general.  Thus,  one might   wonder, why    would  such  quantum gravity effects    not  make  an appearance
     in  ordinary  situations,  perhaps   only as  highly suppressed effects,    associated with radiative corrections involving things  like virtual black holes?

At this  point,   it seems  natural  to  offer the following  speculative   idea (first  described in \cite{Okon2}):  If we accept that   the Hawking    evaporation   of large   black holes leads to  a loss of information and   to the  associated  deviation from quantum mechanical  unitary evolution,  one might wonder   about the  effects of   virtual  black holes  that can be expected to  contribute,  just  as any other quantum  process to  the sum over histories  associated  with a quantum theory that   includes the  space-time  degrees  of freedom.  One  might at  first  argue that  such objects    might have to  correspond to    something  like the Planck scale  mass, and  that  their  corresponding  Compton wavelength  would be  too large  for  the  formation of  a  black hole,  however   after  a  second look, it  seems  unclear    why   should   larger  virtual black holes  be excluded  from  contribution, even if   only  with an   exceedingly small amplitude, and    furthermore,  why   such  classical considerations, regarding for instance  the ratio  of mass  to size for the formation of a black hole,  should  apply to  virtual objects which  can, in principle, be ``off-shell" \cite{George}.  Needless is to say that we  can not expect  clear cut answers to the  above  questions in the  absence of a fully  satisfactory and  workable theory of quantum gravity.
However,  if we  accept that something  that might be characterized,  loosely speaking,   as  virtual black holes,    should contribute,  given the    quantum  nature of  gravity,    to  all  physical process, we would be led to expect, that some  amount of   information loss,  and of  departure of  unitarity  evolution   should be an  intrinsic  part of the effective  description of quantum  evolution  in general.  This, in turn,  seems to    bring  us  to expect a   departure  from Schr\"odinger    evolution, not unlike the one   consider in the modified  theories   involving spontaneous  collapse  such as  GRW  and  CSL.  The  global picture   that  emerges requires adopting  a view about   our theories  of nature that allows   for a kind  of  self referral, which  reminds us of the boot-strap ideas  in  the physics  of strongly interacting particles, and  that  is  discussed in  more detail  in \cite{Okon2}.  We  should note  that, in a sense, this  views  do not seem to  be  very  distant, at least in  spirit,  to   those in    work  \cite{Unruh-Wald}, although  there,   the  connection with the  measurement problem  seems not to have    been made.


     We  should note  here the   early  arguments \cite{Penrose-BH-Collpase}   made  by   R. Penrose   in   connection of  the  possibility of thermal  equilibrium and  ergodic behavior    in
     situations  involving     matter    in   quasi-flat  space-time  and   fully formed  black  holes.  Those   arguments strongly  indicated the need to incorporate  something like the   stochastic
     reduction  of   the quantum state of matter fields in  ordinary conditions, in order to   have a self consistent picture  of such   equilibrium situations.

One  aspect  of the present  proposal that we  find particularly appealing is the fact that,   in contrast  with  most other attempts to  offer a resolution of the BHIP, and which are designed  specifically to  address  just that specific  problem, the  dynamical   collapse theories   were  designed to address  a very different  difficulty in our  current theoretical description of  nature: the measurement problem in quantum mechanics. We find  it truly remarkable that   such proposals  seem to be  able  (when  appropriately refined and modified) to deal  also  with the BHIP.
   Furthermore,   as  was  already  pointed  out in  \cite{Okon2},  the   scheme  we  have  presented  has the   appealing feature of connecting   the   plausible resolution of  other problems  afflicting our understanding of the   workings of nature  in a single unified  picture: the measurement problem,  the emergence of the  primordial inhomogeneities  during inflation \cite{Collapse-and-Inflation}, and the   problem of time  in  canonical quantum theories of gravitation.

    It should  also be pointed  out,  that  part of   the emphasis  in looking for  a resolution of the black  hole evaporation problem,  in a  way that ensures the   unitary    connection  between initial and final states,  comes  from the AdS/ CFT  conjecture.  However,  if the   quantum mechanical  evolution on  the two sides  of the   duality,  is  controlled  by a modified evolution law,  involving non-unitarity and  stochasticity,  such as GRW or CSL, the  duality might still be  possible  even  if the  end     result   of   the evolution   of a black hole that  forms out of a pure state  is  a thermal state.

It  is  clear    that    in    the   course of  the  present  analysis we   have  assumed that a  theory   of quantum gravity would   ``resolve "  the singularity,  and that it leads to   no   gross   violations of  conservation laws  (which would otherwise have   potentially observable   implications)  in   the   regions where   something close to a  classical space-time description is expected to   be  a  very   good  approximation

We  acknowledge that  we made  several  simplifying assumptions  and  that we have  ignored  certain  delicate  and  difficult   issues.  However  we   can  expect   the general picture we obtained to be    rather robust.   We next   deal  briefly  with the most important  of these approximations and issues:


\begin{enumerate}

\item {\it Choice of  collapse  operators}:  The  question of  what   are the  operators $A_{\alpha}$  which   control the CSL  dynamics   as  a unified  and  universal   rule    is an  open issue.  It  is  clear  a complete theory  should specify what determines  such operators,  for all possible circumstances,  and in a manner that depends only on  the  dynamics of the system  in question.   The   early studies  in the  context on   the  non-relativistic  many  particle quantum theory indicate  that   they must correspond  to  a kind of smeared  position operators.    In  section \ref{subsec:operators}  we   took these to be  the   particle number operator  for each of  our  partially  localized modes   in the interior  of black  hole  region.     That  choice  was  done for  calculational  convenience.  On the other hand it  is clear that within  the context of  a fully  covariant   theory these  fundamental operators   should to be locally constructed  from the    quantum fields,  and the number operators   we have used  are clearly non-local. This  might  seem  like a serious concern.  However,   it turns out   that  in  these  theories,  one
can rewrite  the same    CSL evolution   in terms of    various  different  choices of operators. In fact, in   the work \cite{pedro},  a  CSL   evolution  that was in principle   controlled  by  operators associated  with  modes of   definite  wavenumber $\vec  k$,   was shown  to be easily  expressed   in terms of   a CSL evolution  controlled  by  operators associated  with   local  field  operators,
 directly connected  with the quantum field  and its momentum conjugate. Moreover, as it is  well  known from considerations of   EPR situations, the   collapse  of  the state into  eigenvalues of  an operator  associated  with a  certain  space-time region,  has  no  influence  whatsoever in the    effective description in terms of density matrix   for the state  restricted to space-time regions that are causally disconnected (that is the reason one can not use  an EPR pair to send   information in a causal way).
  To see  the point  more clearly  lets focus  on  the case of an EPR pair with correlated spin 1/2 particles,  and  use the  standard Alice  and Bob description.  We  know  that   what exactly is the  operator  measured by Bob has no influence in the   density matrix that describes   the particle still in the hands of Alice,  but which she has yet to subject  to a measurement (that is  the reason  such   non-locallity can not be  used to send  signals faster than light).   The situation  with  the CSL  collapse is just   like that (as it  involves  a  kind of non-locality that can not be used to  send  a causal signals \cite{NLcollpase}). That is,  the exact nature of the quantity controlling the CSL collapse that occurs in the  inside region (that with high curvature) will have no effect in the density matrix that describes the  situation outside the black hole.
  On the other  hand   while  in the case of a  standard  type of  EPR setting,  we  might  analyze  the correlations  between the outcomes of Alice vs. Bob measurements  which  will  depend   on the  quantities  measured   by  each one
   (i.e.  for instance  of the various components of the spin), it is  clear    that  things  change dramatically   in our setting,  because  the state of the subsystem corresponding to Bob (i.e. the matter field in the inside BH Region) will simply disappear.
 In other words, the state  of the   quantum field    in   the region  just after  the would  be singularity or  Quantum Gravity region,   should  correspond to something like  a vacuum state   with    negligible  amounts of   energy and of  information,  an thus  there would be  essentially   no  nontrivial correlations to look for.
Thus, we   can expect  such   robustness (\emph{i.e.},  independence  of the precise   choice of   collapse operators in the   region  III) will   apply to the  density matrix  that describes  of the  quantum field  state   on   $\mathscr{I}^{+}_{R}$.

  \item {\it Choice of  foliation}: We  have performed    the  present     study using a  particular  foliation, which  is adapted to  the fact that,   in our proposal,    the CSL  parameter $\lambda$  depends  on the  scalar curvature $R$.
   However, if the    fundamental   collapse theory   is    fully covariant  as  should   be expected  from a  realistic  collapse theory,    the   evolution of the state  in   the   interaction picture  approach we have  been  using, should   be  describable in  terms of  a  Tomonaga-Schwinger equation\footnote{Recall that  the  Tomonaga -Schwinger equation  $ i \delta \kets{\Psi (\Sigma)}  = {\cal {H}}_I(x)\kets{\Psi (\Sigma)}  \delta \Sigma(x)$    gives the   change in the interaction picture for the state   associated  with the  corresponding    hyper surfaces   $\Sigma'$ and  $\Sigma$,  when the   former is  obtained  from the  latter  by    an infinitesimal   deformation   with  four volume  $\delta \Sigma(x)$  around the  point $x$ in $\Sigma$. We are ignoring  here   the   formal aspects  that indicate that strictly speaking the interaction picture does not exist.}  where, instead  of the   interaction    hamiltonian,  we  would have the corresponding  collapse  theory  density operator.  In  such  setting,  any  specific  physical prediction   would be  foliation  independent. Furthermore,  when  considering a  foliation of   a  region of space-time  lying away  far   from the singularity, the  changes in the  state around any  point,  associated  with the   CSL  type evolution   should  be   describable in  terms  of  regular  local  operators. In that   case these    changes   in the state will  be  very  small,   as  the   CSL-like  parameter  $ \lambda$  will  remain  small   in  regions  where the curvature  is  small, and,  in particular,   these   changes   should not   lead to   anything like ``firewalls" on   the  part of the horizon\footnote{Our   references  to  ``the  horizon"  within the setting where the singularity has  been repaved  by the  ``quantum gravity region",  should be  taken to indicate the   boundary of the past  domain of  dependence  of  said  region.} which  is  far  from the singularity. See  the related comments  at  the  end  of sub-section \ref{Haddamard}.

\item  {\it Relativistic Covariance}:  The  issue  above  clearly illustrates  not only the conceptual importance,  but  the  great  relevance  for the problem at hand,  of   the question of  general covariance of the  full scheme.
We  have   based our  treatment  on  a  non-relativistic    model  of  spontaneous   collapse theory, while a truly    satisfactory proposal to deal with the issue   should  be  based on  a  fully covariant  theory.    Fortunately,   the
 early studies \cite{Pearle-To-Rel, GRW-To-Rel}, and the recent   specific proposals  for      special relativistic  versions of  these  type  of theories \cite{Tumulka-Rel, Bedingam-Rel-1, Bedingam-Rel-2, RelColPearle}  indicate  that  we  should  be   able to  address this  shortcoming in    near  future.   We just  note  again  that  in those  theories    the   interaction picture  evolution    should   be  describable in  terms of  a  Tomonaga-Schwinger   equation  where the   interaction    hamiltonian  is   replaced  by the corresponding  collapse  theory  density operator.  In  that  case
  we  should    be able to carry  the    study, within  those  theories,  of  the present  proposal,   although  it is  clear that the task  would  be  far from trivial.

\item   {\it Energy  conservation}:  
When considering an individual  situation,   rather than  an   ensemble,  the  energy of the initial pulse of matter  will, in general,  not be  exactly equal to that corresponding to the  state   with definite  number of particles $\kets{F}^{ext}_R $ that characterizes the  modified matter content  in the asymptotic  region,  once the    black hole   has  evaporated  completely \footnote{In fact,   this     issue  should  arise    in   all schemes   which,   at the ensemble   level,  are described  by a Lindblad  equation,   which  is  designed to  ensure  energy   conservation.  That is   the  equation   defining the  evolution of the    density matrix  is    similar to  that  in
\eqref{CSL8} and  involves operators $A$ which commute with the hamiltonian,  ensuring that  $\bar {\langle  H \rangle} $ remains   constant in time.  However  in  all the
cases,  such as the  situation at hand,  where this  equation  arises  from an  "unraveling",  capable of  describing  the individual  instances of evolution of a single system \cite{Diosi}, in terms of  some stochastic   variable,  the  question as to  whether  energy is conserved in  each individual case  would become  an open one.  In fact,  general considerations   suggest that  energy  conservation  would be violated, but in  a manner which  might be loosely  associated with the   time  energy uncertainty relation \cite{Time-Energy-U}. }. In this regard  we  note   that CSL,  in general,  leads to small violations of energy conservation,   a  fact that  has led to establish the  most  stringent  bounds  on the parameter  $\lambda$ (although modified  covariant theories might   evade  this problems  altogether.  See  for instance  discussions in  \cite{Bedingam-Rel-1, Bedingam-Rel-2, Unruh-Wald}). Moreover,  we note  that if there is a   small   amount  of energy  remaining inside the  black hole region,    and very close to the singularity,   simply because the positive   and negative energy fluxes do not precisely  cancel  each other,  there would   be  no problem, at least  in principle, if  such  energy were to  be  radiated   after  the singularity.
In such case,  the  resulting picture  would  be one    in which most of the  initial  energy   making the  in-falling pulse of matter   would have    been  radiated in standard  Hawking radiation, and  a  small amount of  energy  would remain to  be radiated  towards infinity  from  the ``quantum gravity  region".

\item   {\it Possibility of  radiation  after  the singularity}:  If, as described  above, there is     some  energy left to  be   emitted  from the  quantum gravity  region, the  corresponding  radiation  might, in principle,   contain  a small amount of  the initial information, and  its  state   could even    be correlated with the  radiation emitted   in the earlier stages.  However,  as  such  energy  will be  minuscule  compared  with the  initial  energy that led to the  formation of the black hole, following  standard  arguments, we do not expect the amount of information that  the  corresponding field  can  encode, to  be significant, thus even  when   contemplating this possibility,  the essence of the picture   we  have presented here,   would remain unchanged.

\item {\it Backreaction}:
  In  the   discussion   so  far, we  have omitted  the   very important  issue of  back reaction.  The  changes  in the  space-time  metric as  a response to those  in the state  of
 the matter  fields,  are   crucial in   accounting    for the decrease of the  mass of the  black hole that  must   accompany  the  Hawking radiation. 
 This  is   essential   for  the   arguments involving  overall  energy  conservation
which,   in turn,  underlay  all  arguments  concerning the  nature of the space-time  that emerges  on the other   side  of the quantum  gravity region  which replaces the  classical   singularity.
 In  a   more realistic  model,   and  as  a result of the the  backreaction,   the   space-time metric,  and the  black hole's mass  would change,  and   so  would   its ``instantaneous  Hawking temperature".
Then the  ``late  time"  radiation would be, in a  sense,   emitted  at a higher temperature  than the ``early times" one,  leading to the runaway effects  that are  associated  with  the expected  explosive  disappearance of the  black hole  itself.  All  such  effects  are  extremely important  to  obtain a  realistic  picture of the entire   history of formation   and  complete  evaporation of a black hole,  but    none  of that   seems  to require  an essential  modification  of the proposal  we have put  forward in this work .

In fact, the   back reaction  is responsible  for the decrease of the Bondi mass   to  a  very small value, as  one  considers the  very late  parts  of $\mathscr{I}^{+}_{R}$ (in  Fig. \ref{folfig}), and that  is  what  allows  us to consider,   matching  our   space-time, at least in the  the asymptotic  region, with  the one   essentially  empty,   and flat  one   expected to  emerge  on the other side of the  quantum gravity region.

\item {\it Reliance  on semi-classical  gravity}:  The issue   discussed   above  led to  us  to reconsider  the  basic  viability of   semi-classical   gravity, the  framework in which  all the present approach is   based. Before engaging in this  discussion  we  should note   that the  alternative to   semi-classical   treatment of gravity,  would  entail the use of some  quantum   theory of gravitation.  The  issue  is,   however,   not only  that  do  we lack,   at  this  point,  any   such  theory  that is   sufficiently developed to offer  us   in  general a  satisfactory   characterization of  quantum  space-time, but   that,  as indicated in the introduction, these theories (at least the  ones   based on a  canonical approach )   suffer from the  so called  ``  problem of time"  as exemplified  by the Wheeler de Witt,  and  the Loop  Quantum Gravity proposals,  a fact   that   impedes  the  presentation of a truly  space-time  description  of any situation. In fact,  the recovery  of  such   a space-time    picture, in  general, is  an open  problem, and  the only   situations in which  that  seems to be  possible  require   some sort of approximated treatment,  that  involves selecting some physical observable to play the role of a clock , and  then   resorting to some  king of   relative  wave-function characterization of the remaining  variables.  It is not clear at all   how  exactly the questions of unitarity and information  should  be addressed in that  context.

 In the  semi-classical    gravity  scheme one treats the  space-time metric  at the classical level,  but uses as  a  source in  Einstein's  equation  the quantum   expectation value of the  energy momentum tensor, and clearly,  this   is  by no means   free of  problems. The  issue  was  considered   in   \cite{Page}, a manuscript that   is often referred  to, as  showing that  semi-classical gravity   is  at odds  with experimental results.  However, it is  not often  emphasized that  such conclusion is only valid  in those contexts  where quantum mechanical evolution   does not  include  any sort of   measurement-related or spontaneous  collapse of the quantum state.  Therefore, it  is   clear  that   such  conclusion   would  not be relevant  for  our  present proposal.  On the  other hand,   it is  clear that if  one   wants  to incorporate  the   instantaneous  (or  even the  continuous)    reduction or  collapse of the  quantum state  of  matter fields,    Einstein's    semi-classical equation  can not be taken  as  fundamental  and  100\% accurate. That   would   be    simply inconsistent: The fact is that  in general   $\nabla^a \langle T_{ab}  \rangle  \not= 0 $  as   the   collapse  of the  quantum state  takes place,  and therefore  $ \langle T_{ab}  \rangle  $   can not  be equated  with $G_{ab}$, which    satisfies  
 $ \nabla^a G_{ab}=0$ identically.
  The point however, is that when viewing   the   metric  description  of  space-time  as just   an effective and approximate  characterization of the  fundamental entity,  in analogy, say,  with the  way the  Navier -Stokes  (NS) equation can be   used  in hydrodynamics, that   basic   problem might    be  bypassed.    That is,   in the same   way  that one  would not expect  the  NS equation to hold   exactly,  or to be  valid  universally,  one  would not take  the  semiclassical Einstein's  equation  as  a fundamental  description. In other words,  just as one  could  easily  envision   situations  where   the NS  would not be  an appropriate  characterization of  the   fluid,  such as when  an ocean  wave  breaks,   or when the fluid  is undergoing a phase transition,  situations    in which  one can expect  important  local departures from the equation,   one   can  expect   something similar   to    happen  with the  semi-classical   Einstein  equation.
   However,    we must emphasize that  this would not   negate   the validity of  the  equation,  at least  as  a good   approximation, for     describing  a  large  class of   regimes  in terms of macroscopic  and  approximate  characterizations.   It  is clear that more precise  characterizations, both in the  treatment  of fluids  and  of   gravitation,  would involve higher  order and more   complicated terms. In fact,  eventually,  as the phenomena in question  become associated  with  a  scale that is closer  to the   natural  scale of the   more fundamental   underlying  theory,   one  would expect the complete  breakdown of such   effective   description.
  The  initial steps in the exploration of  the  formal  adaptation of this approach  to the use of   semi-classical  gravity  in a context  involving  spontaneous  collapse    has  been  developed \cite{Alberto}.  There the issue  has   been explored in detail in   the  cosmological   setting  as part of the  attempts to understand the breakdown  of  the   symmetries of  the  quantum   vacuum  during inflation,    and  the generation via  quantum fluctuations of the primordial inhomogeneities and an-isotropies, an  issue that  has been the focus  of various previous studies \cite{Collapse-and-Inflation}. A brief   review  of that formalism is   presented in  Appendix \ref{ssc}.  One  should note  in this  regard, the  work  \cite{Carlip}  which  considers  in some  detail the  general arguments against  semi-classical  gravity,  and  which  indicates   that  they  are not as  conclusive   as   they are  sometimes  presented.

\end{enumerate}



 One   more thing that is  worth  discussing at this  stage,  is the contrast  in the overall  picture that emerges in the  present approach  with   that  which    would result  from  schemes   in  which information is  preserved  due to   some exotic  quantum gravity  features.

To   clarify this   it is  convenient to recall one of the early   arguments  against  the   information   preserving proposal that  would   rely on   things like Planck  mass  remnants  from the  black hole evaporation. In those proposals   one  would need  the    complete    quantum  state   describing the  Hawking thermal  radiation  and  the    degrees  of freedom of the remnant   to be pure.  We  have of course  no proof that this is  impossible,  however   the resulting picture seems  rather unnatural   simply because  the  internal  degrees of  freedom  of the  remnant, whose  mass  would  be of the order of  $10^{-5}$ gr.  would    have to    be   essentially  equal to the    number of  degrees  of freedom   associated  with  the  radiation,   which  in this case  would  have  the  same  total energy   as the  corresponding collapsing star  (or gas  cloud) that led to the   formation of   the black hole in the first place.  Thus,  the   essential argument   against these  kind of  proposals is that  it requires  us to accept the existence  of objects   with an  incredible  disparity  between  their energy content and  their number of  available degrees of freedom    \footnote{By   available   degrees of freedom for a  subsystem,  we   are  referring, schematically,  to the  collection of   quantum states $\lbrace \kets{\xi_i}  \rbrace $ of the subsystem   in  question  which  could, conceivably,  be  a relevant  part of the  expansion  of the    complete, pure, state of the full system in  the form   $\Sigma_{ij} c_{ij} \kets{\xi_i} \kets{\chi_j} $  where  $\lbrace \kets{\chi_j}  \rbrace $   are the states  characterizing   of the   reminder of the system.}.
 It  is  worth noting therefore that if,  as  suggested in some  proposals,
information  is  fully  preserved,   via    the    escape of   quantum correlations   through the    quantum gravity region  which  resolved the singularity,  we would  have a   very similar situation.
Namely,   in those  schemes, the picture would be  that  long after the black hole  had    evaporated  completely,  the  quantum state describing the  complete   set  of  degrees of freedom  would be  $100\% $  pure,  and yet   its   energy  will  be  essentially   associated with the quasi-thermal  subset (taking into account the   increase of Hawking temperature as the  black hole mass   decreases), characterizing  the  radiation at infinity,  while the    post  quantum gravity region   would have   very small  energy content, and yet it will have   an  exceedingly large number of  degrees of freedom available \cite{Perez}.    Furthermore, even  if we  wanted to consider  a  scheme whereby  information is   lost,  but only   in association   with the    quantum gravity region,  we  might  face  a similar   naturalness problem. That is,   when considering  an  hypersurface that passes  in the  black hole interior   just to the past of the singularity (or more  precisely, the  quantum gravity region) the same issues  would  arise. In other words,   the state of the quantum  field associated  with such hypersurface  would  be  pure,  and yet  its  energy   content   would  be    divided  into
  the  overwhelmingly  large   component    being carried   away towards infinity by the   Hawking radiation and  the  tiny amount  remaining in the
  very low  mass  black hole   that   would  be there at the very late times.  The point again would be that  these two  components   would have to   have  essentially the same number of  available   degrees of freedom  despite   being so    energetically   dissimilar.  In contrast,   in   the proposal   we have considered   here   the   information  is  erased   as the  mass   of the  black hole  decreases   leading to a  more natural    type of  situation where the available  degrees  of freedom   in  a system  is  bound   by  some  monotonic  function of its   total energy.

 The  discussion above   indicates  that the picture  we   have discussed  in this work   should be   rather  robust, however
we reiterate that, at this point, what we have presented must be regarded  as a  toy model.
 A truly  satisfactory and   realistic  version of these ideas  should be  based on a  generalized   theory,  adapted  generically to quantum field theory on curved  spacetimes,  which should  be fully covariant and in a 4  dimensional setting.   However  we  believe that   reasonable   models  with the  basic  features we have discussed  here  do offer  one of the best  hopes for  resolving, the long  standing question  known as  the ``Black Hole Information Loss Paradox".

\section*{Acknowledgments}
We acknowledge useful discussions with Robert Wald, Elias Okon, Philip Pearle, Bernard Kay, George  Matsas, Alejandro Perez and the participants of the meeting ``Haunted Workshop: Who is afraid of Quantum Theory?'', Tepoztl\'an, M{\'e}\-xico, 2013. IP thanks ICN-UNAM for providing visiting facilities during his sabbatical year when part of this work was done. Two of the authors, SKM and LO, are supported by DGAPA postdoctoral fellowships from UNAM.  We acknowledge partial financial support from DGAPA-UNAM projects IN107412 (DS), and CONACyT project 101712 (DS).

\newpage
\appendix
\section{Non-Hadamard behavior in the {\it out} region} \label{Hadamard}
In this section we shall follow the standard Hadamard prescription to prove the non-Hadamard behavior of an arbitrary particle state (including the zero particle state, i.e, vacuum) in the {\it out} quantization defined in the case of evaporating black hole.
\subsection{The two-point function}
We use the discrete basis by working with wave packets. In this basis the modes are orthonormalized unlike the continuous counterpart. The field operator has the following form
\begin{eqnarray}
\hat{f}^{L/R}(x) &=&  \displaystyle\sum_{n,j}\frac{1}{\epsilon^{1/2}}\int_{j\epsilon}^{(j+1)\epsilon} d\omega \left(e^{2\pi i \omega n/ \epsilon}~\hat{b}_{\omega}^{L/R} v_{\omega}^{L/R}+ e^{-2\pi i \omega n/ \epsilon}~ \hat{b}^{L\dag/R\dag}_{\omega}v_{\omega}^{L*/R*} \right.\notag \\
&& \left. + e^{2\pi i \omega n/ \epsilon}~ \hat{\tilde{b}}^{L/R}_{\omega}\tilde{v}_{\omega}^{L/R}+ e^{-2\pi i \omega n/ \epsilon}~ \hat{\tilde{b}}^{L\dag/R\dag}_{\omega}\tilde{v}_{\omega}^{L*/R*}\right)\\
&& = \hat{f}_{1}^{L/R} (x) + \hat{\tilde f}_{2}^{L/R} (\tilde x).
\end{eqnarray}

In the above expression we again used tilde to denote that these modes are defined interior to the black hole {\footnote{Also, just to distinguish we assume all points $x$ belong to the exterior region II (Fig. \ref{cghs}) and $\tilde x$ is defined in the interior (region III).}}. The wave packets defined exterior to black hole, similar to eq.  (\ref{ulr}), are given by
\begin{equation}
v_{jn}^{L/R}=\epsilon^{-1/2}\int_{j\epsilon}^{(j+1)\epsilon}d\omega e^{2\pi i\omega n/\epsilon}v_{\omega}^{L/R},
\end{equation}
where the modes $v_{\omega}^{L/R}$ are given in eq.  (\ref{eq.5}) and eq.  (\ref{eq.6}). Below we provide an explicit calculation only for the right-moving modes since the other left-moving sector is fully analogous.

We calculate the two-point function in an arbitrary state, say, $\kets{\psi}=\kets{F}^{int}\otimes\kets{F}^{ext}$ of the joint basis. The two-point function takes the following form
\begin{eqnarray}
\bras{\psi} \hat{f}^R(x)  \hat{f}^R(x') \kets{\psi} = \bras{F}^{ext} \hat{f}_{1}^R(x)  \hat{f}_{1}^R (x') \kets{F}^{ext} + \bras{F}^{int} \hat{\tilde f}_{2}^R (\tilde x)  \hat{\tilde f}_{2}^R (\tilde x') \kets{F}^{int}
\end{eqnarray}
Note that the contributions from the exterior (region II, Fig. \ref{cghs}) and interior (region III, Fig. \ref{cghs}) of the black hole decouples from one another. Let us first focus on the first term defined exterior. Calculations involving the other term defined in the interior to the black hole is exactly similar. Writing this term explicitly yields
\begin{eqnarray}
\bras{F}^{ext} \hat{f}_{1}^R(x)  \hat{f}_{1}^R (x') \kets{F}^{ext} =  \bras{F}^{ext} \displaystyle\sum_{n,j} \displaystyle\sum_{n',j'} \frac{1}{\epsilon} \int_{j\epsilon}^{(j+1)\epsilon}d\omega \int_{j'\epsilon}^{(j'+1)\epsilon}d\omega' \notag \\
\left(e^{2\pi i (n \omega - n' \omega')/ \epsilon}~v_{\omega} v_{\omega'}^{*} \hat{b}_{nj}{{\hat b}^{\dagger}}_{n'j'} + e^{-2\pi i (n \omega - n' \omega')/ \epsilon}~v_{\omega'} v_{\omega}^{*} \hat{b}^{\dagger}_{nj}\hat{b}_{n'j'}\right) \kets{F}^{ext}
\label{tpf}
\end{eqnarray}
while the other terms do not contribute. The next step would be to use the commutator relation between the creation and annihilation operators and orthonormality condition
$\bras{F_{nj}}{F_{n'j'}}\rangle^{ext} = \delta_{nn'}\delta_{jj'}$. This simplifies eq.  (\ref{tpf}) to the following form
\begin{equation}
\bras{F}^{ext} \hat{f}_{1}^R (x)  \hat{f}_{1}^R (x') \kets{F}^{ext} =  \displaystyle\sum_{n,j} \frac{(2F_{nj} + 1)}{\epsilon}~I_{1}(\omega) I_{1}^{*}(\omega')
\label{tpf2}
\end{equation}
where,
\begin{eqnarray}
 I_{1} (\omega) = \int_{j\epsilon}^{(j+1)\epsilon}\frac{d\omega}{\sqrt{4\pi\omega}}  e^{\frac{i\omega}{\Lambda}\log(-\Lambda x)+ 2\pi i n \omega/\epsilon} \\
 I_{1}^{*} (\omega') = \int_{j\epsilon}^{(j+1)\epsilon}\frac{d\omega'}{\sqrt{4\pi\omega'}}  e^{\frac{-i\omega'}{\Lambda}\log(-\Lambda x)- 2\pi i n \omega'/\epsilon}.
\end{eqnarray}
In the expression eq.  (\ref{tpf2}), the integration over the term that includes $F_{nj}$ gives
\begin{equation}
 \displaystyle\sum_{n,j} \frac{F_{nj}}{\epsilon(\frac{1}{\Lambda}\log(-\Lambda x)+ \frac{2\pi n}{\epsilon})^{1/2}(\frac{1}{\Lambda}\log(-\Lambda x')+ \frac{2\pi n}{\epsilon})^{1/2}}~\left(\gamma[\frac{1}{2},(j+1)\epsilon]-\gamma[\frac{1}{2},j\epsilon]\right)^2
\end{equation}
Note that since $\kets{F}^{ext}$ is a well defined state in the out Fock space it may have arbitrary number of particles but in no case it can be infinity. Therefore not all $F_{nj}$-s are nonzero
when $n$ and $j$ are changing their values. On the other hand the term without $F_{nj}$ in eq.  (\ref{tpf2}) can be simplified in the following manner. We first use the identity
\begin{equation}
 \displaystyle\sum_{n} e^{2\pi i n (\omega-\omega')/ \epsilon} = \epsilon \delta(\omega - \omega').
\end{equation}
As a result the  integrations of the two  integrals over $\omega$  and  $\omega'$ transforms into  a single integral, and  then,  by appropriately using  the  definition of the function $\gamma$     to turn  the the sum over $j$  into an  integral,   we  obtain the final expression for the two-point function
\begin{eqnarray}
\bras{F}^{ext} \hat{f}_{1}^R (x)  \hat{f}_{1}^R (x') \kets{F}^{ext} &=& - \frac{1}{4\pi} \log|\log(x/x')| \notag \\
&& + \displaystyle\sum_{n,j} \frac{F_{nj}}{2\pi\epsilon h(x) h(x')} \left(\gamma[\frac{1}{2},(j+1)\epsilon]-\gamma[\frac{1}{2},j\epsilon]\right)^2,\notag\\
&& h(x) = (\frac{1}{\Lambda}\log(-\Lambda x)+ \frac{2\pi n}{\epsilon})^{1/2}.
\label{g1}
\end{eqnarray}
In order to check the Hadamard behavior, as explained in  Section 3.3, we need to construct the symmetrized two-point function
\begin{equation}
G^{(1)} (x, x') = \frac{1}{2}(\bras{F}^{ext} \hat{f}_{1}(x)  \hat{f}_{1} (x') \kets{F}^{ext} + \bras{F}^{ext} \hat{f}_{1}(x')  \hat{f}_{1} (x) \kets{F}^{ext})
\end{equation}
and subtract the ``Hadamard ansatz''. If the quantum state is Hadamard then we expect no singular behavior after the above said subtraction.

\subsection{The Hadamard Ansatz}

In two dimensions it seems to be a non-agreement about precise
 form of the ``Hadamard ansatz" \cite{yDecaFol08} and \cite{hSalyBi01}. Usually the differences in opinion comes through the finite terms that are present in the Hadamard ansatz. Nevertheless, for our purpose we are interested to extract the singular behavior due to coincidence limit. We restrict to the following singular behavior of the Hadamard ansatz
\begin{eqnarray}
H (x, x') = -\frac{1}{4\pi} \log\sigma,
\label{hada}
\end{eqnarray}
where $\sigma$ is half of the geodesic distance square between two close points $x$ and $x'$ in the normal neighborhood.

Now let us proceed to calculate the geodesic distance between $x$ and $x'$. We will do it in null coordinates. The calculations of this section are based on section 2 of chapter II of \cite{jlSyn71}. The geodesic distance is given by
\begin{equation}
\sigma=\frac{1}{2}u_{1}^{2}g_{\mu'\nu'}U^{\mu'}U^{\nu'},
\label{eq:2}
\end{equation}
where $u_{1}$ is the affine parameter of the geodesic at $x'$, $U^{\mu'}=\frac{dx}{du}$ is the tangent vector to the geodesic and $g_{\mu'\nu'}$ is the metric of space-time.
Now we use that
\begin{equation}\label{eq:3}
u_{1}U^{\mu'}=\xi^{\mu}+\frac{1}{2}\Gamma^{\mu'}_{\alpha'\beta'}\xi^{\alpha}\xi^{\beta},
\end{equation}
where $\xi^{\mu}=x^{\mu}-x^{\mu'}$ and $\Gamma^{\mu'}_{\alpha'\beta'}$ are the usual Christoffel symbols. By using eq.  (\ref{eq:3}) in eq.  (\ref{eq:2}) we obtain
\begin{equation}
\sigma=\frac{1}{2}g_{\mu'\nu'}(\xi^{\mu}+\frac{1}{2}\Gamma^{\mu'}_{\alpha'\beta'}\xi^{\alpha}\xi^{\beta})(\xi^{\nu}+\frac{1}{2}\Gamma^{\nu'}_{\gamma'\eta'}\xi^{\gamma}\xi^{\eta}).
\end{equation}
Up to cubic order in $\xi^{\mu}$ we obtain
\begin{equation}\label{eq:4}
\sigma=\frac{1}{2}(g_{\mu'\nu'}\xi^{\mu}\xi^{\nu}+\frac{1}{2}g_{\mu'\nu'}\Gamma^{\mu'}_{\alpha'\beta'}\xi^{\alpha}\xi^{\beta}\xi^{\nu}+\frac{1}{2}g_{\mu'\nu'}\Gamma^{\nu'}_{\gamma'\eta'}\xi^{\mu}\xi^{\gamma}\xi^{\eta}+ ...).
\end{equation}
Now we find that the only Christoffel symbols different from zero are {\footnote{Here we use the redefinition $x=x^- + \Delta.$}}
\begin{equation}\label{eq:5}
\Gamma^{x}_{xx}=x^{+}\Omega\Lambda^{2},\hspace{0.2cm}\Gamma^{x^{+}}_{x^{+}x^{+}}=x\Omega\Lambda^{2},
\end{equation}
where $\Omega=\frac{1}{\frac{M}{\Lambda}-xx^{+}\Lambda^{2}}$ and $g_{xx^{+}}=g_{x^{+}x}=-\frac{1}{2}\Omega$. Then the geodesic distance is given by
\begin{eqnarray}
\sigma &= &\frac{1}{2}(2g_{x'x'^{+}}\xi^{x}\xi^{x^{+}}+\frac{1}{2}g_{x'x'^{+}}\Gamma^{x}_{xx}\xi^{x}\xi^{x}\xi^{x^{+}}+\frac{1}{2}g_{x'x'^{+}}\Gamma^{x'^{+}}_{x'^{+}x'^{+}}\xi^{x^{+}}\xi^{x^{+}}\xi^{x} \\\nonumber
& + & \frac{1}{2}g_{x'x'^{+}}\Gamma^{x'^{+}}_{x'^{+}x'^{+}}\xi^{x}\xi^{x^{+}}\xi^{x^{+}}+\frac{1}{2}g_{x'^{+}x'}\Gamma^{x'}_{x'x'}\xi^{x^{+}}\xi^{x}\xi^{x}).
\end{eqnarray}
This simplifies to
\begin{equation}\label{eq:6}
\sigma=g_{x'x'^{+}}\xi^{x}\xi^{x^{+}}(1+\frac{1}{2}\Gamma^{x'}_{x'x'}\xi^{x}+\frac{1}{2}\Gamma^{x'^{+}}_{x'^{+}x'^{+}}\xi^{x^{+}}).
\end{equation}
Using eq.  (\ref{eq:5}) and the expression for the metric in eq.  (\ref{eq:6}) we finally obtain
\begin{equation}
\sigma=-\frac{1}{2}\Omega\Delta x\Delta x^{+}(1-\frac{\Omega\Lambda^{2}}{2}(x^{+}\Delta x+x\Delta x^{+})),
\label{geo}
\end{equation}
where $\Delta x=x'-x$ and $\Delta x^{+}=x'^{+}-x^{+}$. This geodesic distance is clearly an approximation, but for our purposes is seems to be enough.

\subsection{The non-Hadamard behavior}

In order to check the Hadamard property we substitute eq.  (\ref{geo}) in eq.  (\ref{hada}) and subtract the result from eq.  (\ref{g1}). We obtain the following expression
\begin{eqnarray}
F(x, x') &&= - \frac{1}{8\pi} \log(xx') - \log[-\frac{1}{2}\Omega\Delta x^{+}(1-\frac{\Omega\Lambda^{2}}{2}(x^{+}\Delta x+x\Delta x^{+}))]\notag\\
&& + \displaystyle\sum_{n,j} \frac{F_{nj}}{2\pi\epsilon h(x) h(x')} \left(\gamma[\frac{1}{2},(j+1)\epsilon]-\gamma[\frac{1}{2},j\epsilon]\right)^2.
\end{eqnarray}
Note that with the above expression for the renormalized two-point function we have a well defined coincidence limit with $x$ and $x'$. However while approaching the horizon $x= x^- + \Delta =0$, there appears another divergence of the form $\log x$ which is not of the Hadamard form. Therefore clearly the state $|F\rangle^{ext}$ is not Hadamard. Similarly one can prove that the particle state $|F\rangle^{int}$ is also non-Hadamard. This obviously implies that the states $|F\rangle^{int}\otimes|F\rangle^{ext}$ of the joint Hilbert space is non-Hadamard.

\section{Proof of well defined foliation} \label{proof}

The intersecting curves eq.  (\ref{txin}) and eq.  (\ref{xtout}) are very important since they determine the foliation of the space-time. One might be concerned with the possibility that the foliating surfaces may cross each other at some region of the space-time. If this happens then one fails to associate a well defined evolution of the quantum states. Here we give a proof that such crossings among foliating surfaces does not take place.

\begin{figure}
\centering
\includegraphics[scale=0.31]{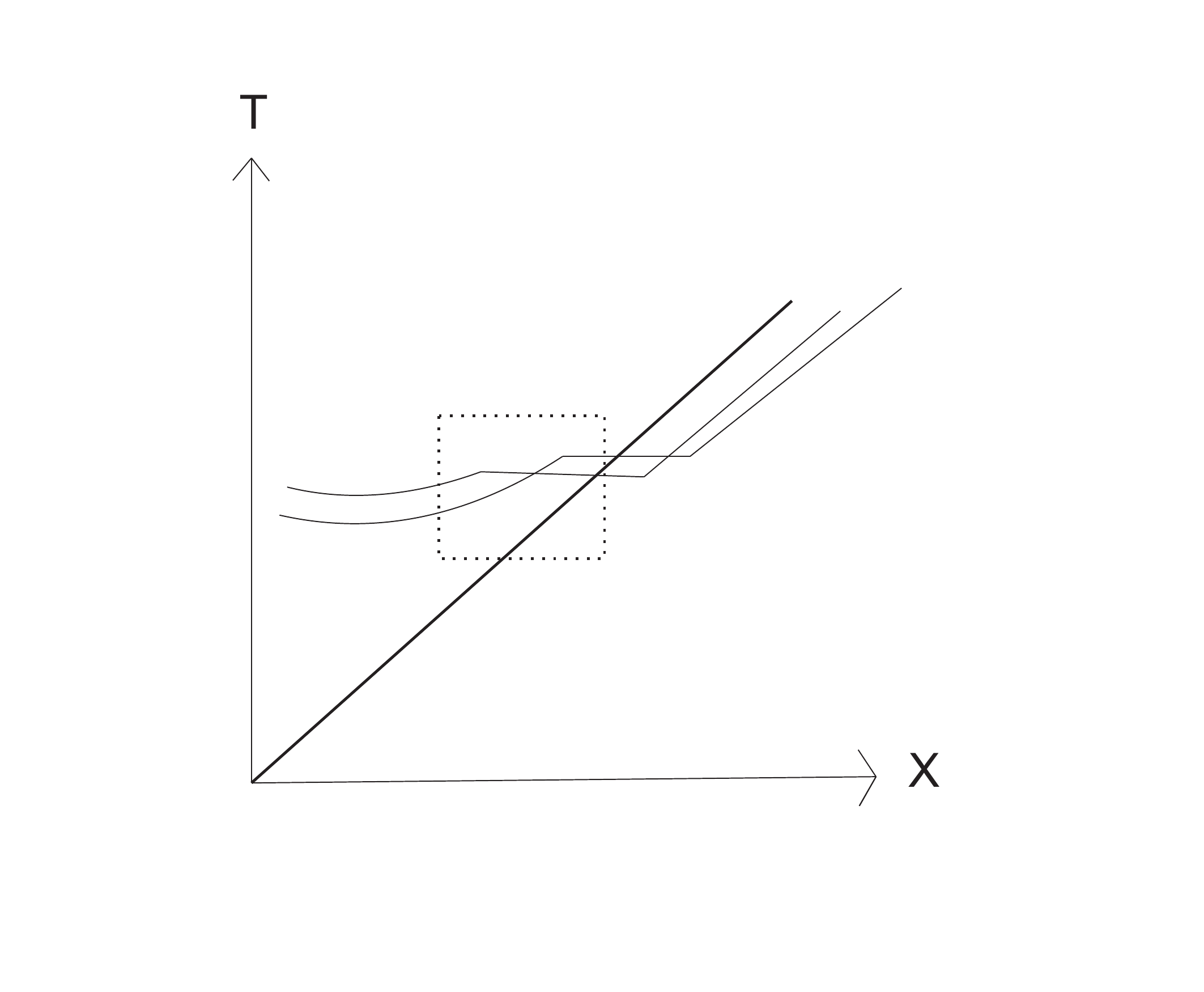}
\includegraphics[scale=0.31]{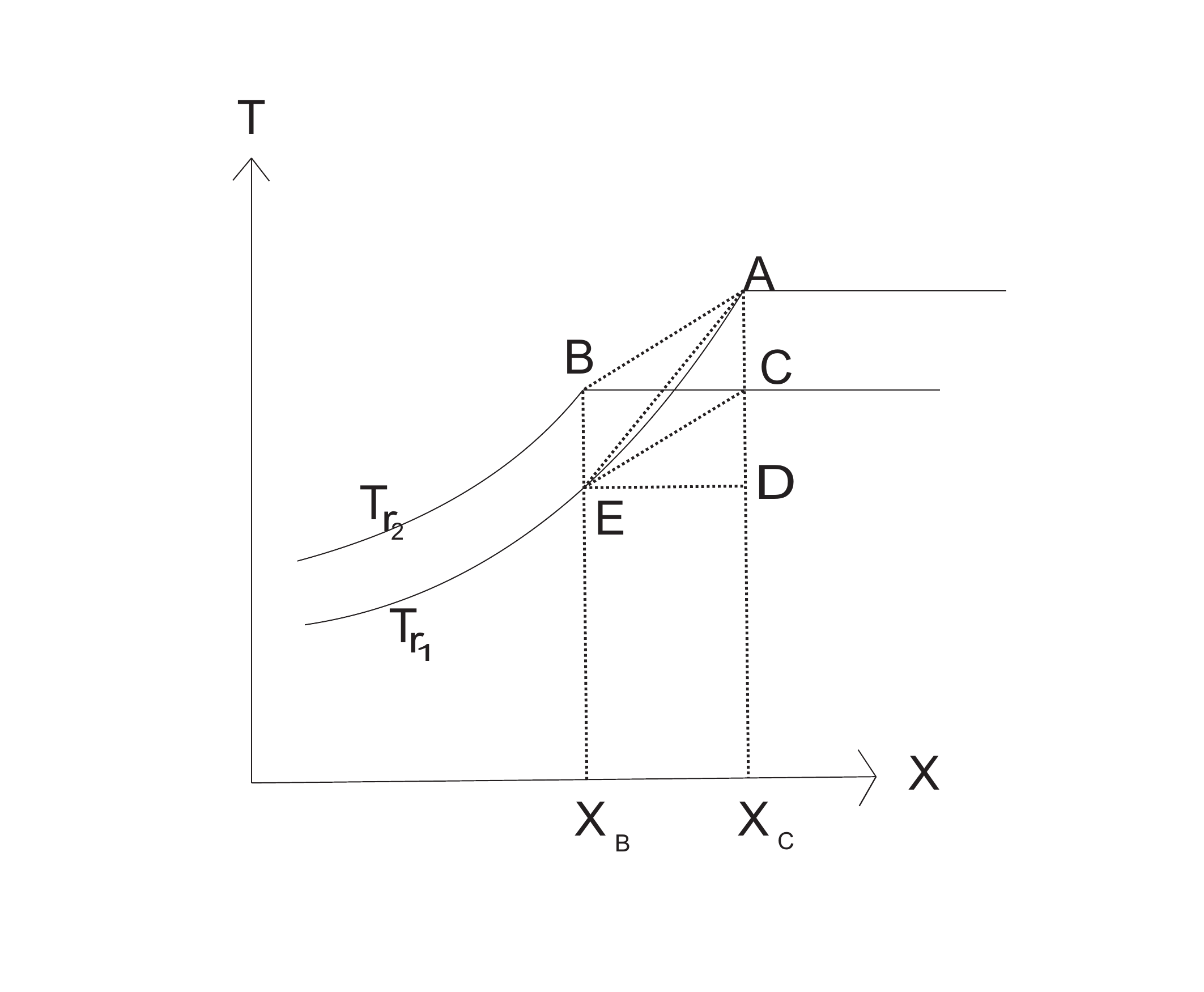}
\caption{Possible crossing between hyper-surfaces due to a bad choice of foliation. We show this situation does not take place in our construction. For details see the text.}
\label{folproof}
\end{figure}

The most general situation where this crossing can happen is shown on the left in Fig. \ref{folproof}. From the figure we can see that such crossings can take place both in regions II and III. Let us first consider the situation in region III. A zoomed version highlighting this aspect is shown on the right of Fig. \ref{folproof}. In the figure $T_{r_1} (r_1, X)$ and $T_{r_2}(r_2, X)$ are two $r=const.$ curves and the points $A$ and $B$ are on curves $T_{r_1}$ and $T_{r_2}$ respectively. The slope of the line joining $A$ and $B$ is given by
\begin{equation}
m_{BA} = \frac{T_{A} - T_{C}}{X_C - X _B}.
\end{equation}
This can be easily expressed in terms of the slopes of the line joining the points $E$ with $A$ and $E$ with $C$ such that
\begin{equation}
m_{BA} = m_{EA} - m_{EC}.
\label{slopes}
\end{equation}
Now making use of the mean-value theorem and $\Delta X = X_C - X _B \rightarrow 0$ we can associate the slope $m_{BA}$ with the slope of the guiding curve (generated by all intersecting points) and $m_{EA}$ with the curve $T_{r_1}$. Then eq.  (\ref{slopes}) clearly tells us that the slope of the guiding curve should be less than the slope of $r=const.$ curve. But this is clearly in contradiction with our choice eq.  (\ref{txin}). Actually, it is trivial to verify that the slope of eq.  (\ref{txin}) is always greater than the slope of eq.  (\ref{rin}) provided $r<r_h$. With this we rule out the crossing behavior showing in Fig. \ref{folproof}. Similarly, for region-II, in order to have a crossing it is necessary that the slope of the guiding curve is less that the $t= const.$ curve. However, by comparing the slopes of eq.  (\ref{xtout}) and eq.  (\ref{tout}) one can conclude that this situation does not appear given the condition $X>T$ in region II.

\section[Useful integrals]{Useful integrals to define $\zeta$} \label{int}
Here we provide the explicit results of various integrals defined in sub-section \ref{sec:appl_CSL}. These integrals measure the distance of a point in a particular foliating hypersurface. This distance is the parameter $\zeta$ eq.  (\ref{distance}) used together with $\tau$ for locating a point in CGHS space-time.
\begin{eqnarray}
\zeta_{A} &=& \left(\frac{A}{M/\Lambda - \Lambda^2 A}\right)^{1/2} \ln|\frac{c-e}{f}| \notag\\ &&A = \frac{e^{2\Lambda r (c,e)}}{\Lambda^2}, \nonumber
\end{eqnarray}

\begin{eqnarray}
\zeta_{B} &=&\frac{1}{\Lambda} \log \left(-2 c \Lambda \right. \nonumber \\
&& \left. +2 \sqrt{\frac{-2 c \Delta  \Lambda ^3-2 c \Lambda ^3 x+M+\Lambda ^3 x^2+\Delta  \Lambda ^3 x}{\Lambda }}+\Delta  \Lambda +2 \Lambda  x\right)_{c-e-\Delta}^{c-d-\Delta} \nonumber
\end{eqnarray}

\begin{eqnarray}
\zeta_{C} &=& \frac{1}{\Lambda}\log \left(\Delta  \Lambda +2 \Lambda  x \right. \nonumber \\
&& \left. + 2 \sqrt{\frac{-a M+a \Lambda ^3 x^2+a \Delta  \Lambda ^3 x+b M+b \Lambda ^3 x^2+b \Delta  \Lambda ^3 x}{\Lambda  (a+b)}}\right)_{c-d-\Delta}^{a-b}. \nonumber
\end{eqnarray}

\section{The SSC  Formalism and the  Sudden  Collapse of the  quantum state}
\label{ssc}

When one  considers   a   description of  a  situation requiring  a quantum  treatment of matter
fields  and at the same time   requires  a classical  picture of space-time, allowing the
consideration of questions  such as those that   usually   as  arise in the   context of the   BHIP;  one is  in  the realm of     semi-classical  gravity.  As  discussed in   this can   not
 considered as  a   truly fundamental   characterization  of  the physical situation ,  but   only as  an effective  description.  Nevertheless it is  convenient  among   other reasons  for    the sake of  conceptual  clarity  to   have a  well defined   framework,  where various issues can be  studied    in a self consistent  manner.  This  can be  thought  as  analogous to,  say, the Navier Stokes
 equation in    hydrodynamics, which   even though can not be  considered as a fundamental
 description of a fluid,   and  which   is known not to be  valid under various   kind  of
 circumstances,     leads to  an internally self consistent  description, that can be the subject of
 rigorous  mathematical  analysis.  A scheme  which is  hoped to  offer a   similar characterization
 has  been developed  for   semi-classical gravity,  and it is    defied  as follows.

{\bf Definition :}
The set $\left\lbrace g_{\mu\nu}(x),\hat{\varphi}(x),\hat{\pi}(x),\mathscr{H},\vert\xi\rangle\ in \mathscr{H}\right\rbrace$
represents a
\textit{Semiclassical Self-Consistent Configuration}
SSC if and only if $\hat{\varphi}(x)$, $\hat{\pi}(x)$ and $\mathscr{H}$  correspond to  a quantum field theory
constructed over a space-time  with  metric  $g_{\mu\nu}(x)$ (as  described in, say \cite{Wald}), and  the  state
$\vert\xi\rangle$ in $\mathscr{H}$  is  such that
\begin{equation}\label{Mset-up}
G_{\mu\nu}[g(x)]=8\pi G\langle\xi\vert \hat{T}_{\mu\nu}[g(x),\hat{\varphi}(x),\hat{\pi}(x)]\vert\xi\rangle
\end{equation}
for all the points $x
$ in the space-time manifold.

\vspace{0.3cm}
  The  point however is that  such  rigid  definition does not   allow  for the incorporation of
  something like  a   collapse of the quantum state  of the matter  fields.  However   we  can
  extend the scheme   by  incorporating a  transition from  one  SSC  to  a  another  associated to a
  sudden  change in the state of the  system, as an  analog  of the  matching  used  in  the
  treatment of  thin shells  in  general relativity  developed in \cite{Israel}. That is,  just as in the
  later  case,  one  matches two  space-times across a  time-like boundary representing the thin
  matter  shell,  in the   situation at hand,  one matches   two space-times   across  a space-like
  hypesurface taken to represent  the   excitation of the fundamental underlying  DOF  of   quantum
  gravity that must occur in association  with what we call the collapse of the  quantum state of
  the matter fields.

  The  basic idea  is then that a collapse is  associated  not just  with  the  sudden transition  from
  one state $\vert\xi_1\rangle$ in $\mathscr{H}$ to  another $\vert\xi_2\rangle$ in $
  \mathscr{H}$   but  with a  the transition of a    complete SSC  to  another $ SSC_1 \to  SSC_2$,
  which not only involves a jump in the  state  but also the jump in the   space-time metric  and
  even  in the   Hilbert space.  The analysis    performed in \cite{Alberto}  indicates that  the
  matching of space-times might be  done   while requiring continuity of the  induced    metric
  across  the   collapse hypersurfae,  but  allowing   for a discountinuity  of the   extrinsic
  curvature.

   The  above  scheme  is    clearly  designed to deal with a single  spontaneous   collapse of the
   quantum state, but   it seems in principle extendable  to  deal with proposals   involving
   continuous   reduction processes  such as   those  considered  in  CSL.

     It is  clear  however that further  research on the  details of    these  formalism  as  applied to   theories like  CSL  is   required.

\end{document}